# Visualizing electron correlation by means of ab-initio scanning tunneling spectroscopy images of single molecules

Dimitrios Toroz, Massimo Rontani,* Stefano Corni

*Centro S3, CNR – Istituto di Nanoscienze, Via Campi 213/a, 41125 Modena, Italy*

Scanning tunneling microscopy (STM) has been a fundamental tool to characterize many-body effects in condensed matter systems, from extended solids to quantum dots. STM of molecules decoupled from the supporting conductive substrate has the potential to extend STM characterization of many body effects to the molecular world as well. In this article, we describe a many-body tunneling theory for molecules decoupled from the STM substrate, and we report on the use of standard quantum chemical methods to calculate the quantities necessary to provide the 'correlated' STM molecular image.

The developed approach has been applied to eighteen different molecules, to explore the effects of their chemical nature and of their substituents, as well as to verify the possible contribution by transition metal centers. Whereas the bulk of calculations have been performed with CISD because of the computational cost, some tests have been also performed with the more accurate CCSD method to quantify the importance of the computational level on many-body STM images.

We have found that correlation induces a remarkable squeezing of the images, and that correlated images are not derived from Hartree-Fock HOMO or LUMO alone, but include contributions from other orbitals as well. Although correlation effects are too small to be resolved by present STM experiments for the studied molecules, our results provide hints for seeking out other species with larger, and possibly experimentally detectable, correlation effects.

* Corresponding author. E-mail: rontani@unimore.it

## I. Introduction

Scanning tunneling microscopy and spectroscopy (STM and STS) is a fundamental tool for the study of the intimate property of matter at the atomic scale.[1] In particular, STM has been pivotal in characterizing many-body (i.e., electron-electron correlation) effects in condensed matter systems. Just to mention a few examples, STM has been applied to investigate the *d*-wave pairing in high-$T_c$ superconductors,[2] the many flavors of magnetic order,[3] the Luttinger liquid state in metallic carbon nanotubes,[4] the quasi-particle lifetime in graphene,[5] exotic types of electron crystallization,[6] the Kondo resonance between substrate and either magnetic impurities[7] or surface states.[8]

With regards to zero dimensional systems, recent studies have shown both experimentally[9-11] and computationally[12-15] that the STS images in space of electronic states of semiconductor quantum dots may be dramatically different from what predicted by the standard mean-field model, as manifestations of strong electron correlation in the dot. The amount of distortion of the STS images has been found to depend on the degree of electronic correlation. We stress that this effect: (i) it is related to Coulomb blockade physics;[16] (ii) it is inherent in the few-electron zero dimensional system; (iii) it is distinct from other many-body effects observed in tunneling spectroscopies of quantum dots, such as the Kondo effect[17] and the Fermi edge singularity.[18] The latter descend from the significant coupling between dot and electrodes (due to either tunneling or Coulomb interaction).

Considering the conceptual similarity of quantum dots with atoms and molecules, the natural following step is to investigate whether STS images of *molecules* may be modified by electronic correlation. This study is timely since Coulomb blockade has already been demonstrated in single molecules.[19] As in the case of quantum dots, the molecule should be decoupled from the underlying conductive substrate, a condition that has been recently achieved in STM experiments either by inserting an insulating layer between molecule and conductive substrate[20-23] or by considering physisorbed molecules.[24] In fact, the Coulomb blockade physics that lays at the basis of the tunneling experiments through isolated zero-dimensional systems (quantum dots or molecules) substantially differs from the case of extended systems. For the latter, the electron transfer between the STS tip and the delocalized electronic states does not perturb substantially the system, whereas the ionization, positive or negative, of a molecule affects its energy and wave function, due to both the confinement effects on the nanometric length-scale and the electron-electron interaction.

Whereas several approaches have already been proposed to simulate STS images of molecules,[25-34] we are aware of only one work where correlation effects on STS images have been investigated.[30] In this article, Nakatsuji and coworkers used a discrete model of the tip + substrate system to calculate STM images of $Li_2$ taken with a $Li_2$ tip at the CI and SAC/CI level, and concluded that in this system electron correlation appreciably affects STM matrix elements.

In the present paper, we describe an extension of the many body tunneling theory previously developed for quantum dots[9,10,12,13] to molecules. At odds with Ref. 30, a discrete model of the system including the tip is not assumed *a priori*, which allows us to focus on correlation effects intrinsic of the probed molecule. Moreover, we describe here how quantum chemical methods can be used to calculate the quantities that appear in the developed theory, so to provide the ‘correlated’ STS image for the single molecule.

It is to be stressed that our aim is not to propose yet another method to simulate and interpret experimental STS images. Our aim is instead to show how the electronic correlation intrinsic of the molecule can be revealed in terms of distortion of STS images, providing a theoretical and computational apparatus to quantify and visualize such distortions. Correlation effects (although mostly related with the molecule-conductive substrate interactions) have been already demonstrated for the electronic energy levels of the adsorbed molecule.[35]
The protocol we have used allows for computing the many-body spectral density (the quantity that is directly related to STS image, see Sect. II) from the results of a widespread computer code (Gaussian03)[36] and visualizing it with the aid of a commonly used graphical tool (VMD).[37]

The quantum chemical method we have chosen in this study to perform the bulk of calculations is the configuration interaction method with single and double excitations (CISD). CISD is the minimal non-perturbative correlation method, thus we compromise between accuracy and feasibility of the calculations. To check the result of higher correlated methods we have performed selected tests with the coupled cluster with single and double excitations (CCSD) method.

We have applied the developed approach to eighteen molecules, including the experimentally studied pentacene.[20] We have mostly focused on planar aromatic molecules, chosen to explore the effects of the size of the conjugation as well as of the substituent electronegativity, although we have also considered a simple metal complex. The results of these tests show that the effect of electron correlation on the STS images of the frontier orbitals is two-fold: (i) It squeezes the orbitals, with consequent loss of spectral weight. (ii) It hybridizes the orbitals, mixing them with virtual and occupied states consistently with the symmetry group of the molecule. These effects are difficult to be observed directly with present STS resolution. However, a few hints to guide further investigations toward molecules with more visible correlation effects may be grasped from the results (see Sect. IV).

The plan of the rest of the paper is as follows: After the theoretical analysis of the quasi-particle wave function (Section II), we illustrate the computational method to create the ab-initio STM images (Section III). Then we discuss the effect of correlation on a series of tested molecules (Section IV) and finally we summarize our findings in Section V. Supplemental material is attached at the end of the main text.

## II. Quasi-particle wave function

For the sake of clarity, here we recall the theory of the *quasi-particle* wave function (QPWF) developed in References 12 and 13 in the context of semiconductor quantum dots. In this section we illustrate its straightforward extension to the case of single molecules.

The space- and frequency-resolved spectral density $N(\mathbf{r},\omega)$ is the many-body observable ideally accessible by low-temperature STS experiments.[9,10,12,13,38] The explicit expression of $N(\mathbf{r},\omega)$ for hole-like excitations ($\hbar\omega < \mu$) is:

$$N(\mathbf{r},\omega) = \hbar\sum_i \left|\left\langle N-1,i\right|\hat{\Psi}(\mathbf{r})\left|N,0\right\rangle\right|^2 \delta(\hbar\omega - E_0(N) + E_i(N-1)), \qquad (1)$$

where $\left|N,i\right\rangle$ is the many-body *i*th excited state of the molecule + substrate system with *N* electrons whose energy is $E_i(N)$ (*i* = 0 for the ground state), $\mu$ is the chemical potential, $\hat{\Psi}(\mathbf{r})$ is the Fermi field annihilation operator destroying an electron at position $\mathbf{r}$. If the spacing between hole-like excitation energies $E_0(N) - E_i(N-1)$ close to $\mu$ is larger than the available energy resolution $\hbar d\omega$, then integrating Eq. (1) over the neighbor of $\mu$ one obtains $N(\mathbf{r},\omega)d\omega = \left|\left\langle N-1\right|\hat{\Psi}(\mathbf{r})\left|N\right\rangle\right|^2$ (putting $\left|N\right\rangle \equiv \left|N,0\right\rangle$). This (positive-definite) spectral density may be regarded as the square modulus $\left|\varphi(\mathbf{r})\right|^2$ of the QPWF which has been subtracted (hole QPWF, h-QPWF, transition from *N* to *N* – 1 electrons) to the *N*-body system:[12,13,41]

$$\left|\varphi(\mathbf{r})\right|^2 = \left|\left\langle N-1\right|\hat{\Psi}(\mathbf{r})\left|N\right\rangle\right|^2. \qquad (2)$$

A similar definition holds for the QPWF added to the system (electron QPWF, e-QPWF, transition from *N* to *N* + 1 electrons): $\left|\varphi(\mathbf{r})\right|^2 = \left|\left\langle N+1\right|\hat{\Psi}^+(\mathbf{r})\left|N\right\rangle\right|^2$.

The idea that tunneling spectroscopy is a sensitive probe of the quasi-particle excitations of the many-body system has first been put forward in the context of superconductors.[42] Recently it has been shown both theoretically[12-15] and experimentally[9-11] that STS (magnetotunneling) imaging of electron states in semiconductor quantum dots may provide the map of $\left|\varphi(\mathbf{r})\right|^2$ in real (reciprocal) space. In particular, this connection has been explicitly illustrated by Rontani and Molinari[12] for quasi two-dimensional quantum dots in the framework of the effective mass approximation. On the other hand, this relation is general and holds for STS spectroscopy of molecules and other nano-objects as well. In Discussion S1 of the Supplemental Material we confirm this by working out an explicit expression for the STS differential conductance in terms of the QPWF of the molecule under the assumption of a locally spherical tip.[43]

The spectral density $N(\mathbf{r},\omega)$ may be regarded as the many-body generalization of the density of states $\rho(\mathbf{r},\omega)$, which is routinely computed within the framework of the mean-field treatments of Coulomb interaction, such as Hartree-Fock and practical implementations of density functional theory. The density of states $\rho(\mathbf{r},\omega)$ is a sum over occupied single-electron (Hartree-Fock or Kohn-Sham) spin-orbitals, whose orbital wave functions are $\phi_\alpha(\mathbf{r})$ and energies $\varepsilon_\alpha$, with

$$\rho(\mathbf{r},\omega) = \hbar \sum_\alpha \left|\phi_\alpha(\mathbf{r})\right|^2 \delta(\hbar\omega - \varepsilon_\alpha) \quad (3)$$

and $\alpha$ being the set of pertinent quantum indexes, including spin. If the spacing between energy levels $\varepsilon_\alpha$ is larger than the available energy resolution $\hbar d\omega$, then $\rho(\mathbf{r},\omega)d\omega$ reduces to the square modulus of a single orbital, $\rho(\mathbf{r},\omega)d\omega = \left|\phi_\alpha(\mathbf{r})\right|^2$.

Note that Hartree-Fock theory neglects correlations beyond mean field and density functional theory considerably simplifies them. On the other hand, the spectral density $N(\mathbf{r},\omega)$ takes all many-body correlations into account. In particular, the quasi-particle is the extra particle dressed by the interaction with the other $N$ electrons of the system. As we show next, in the absence of correlation the e-QPWF (h-QPWF) reduces to the Hartree-Fock LUMO (HOMO), $\varphi(\mathbf{r}) \to \phi_{\overline{\alpha}}(\mathbf{r})$, where $\overline{\alpha}$ is the index labeling the frontier orbital.

For any practical calculation of

$$\varphi(\mathbf{r}) = \left\langle N-1 \middle| \Psi(\mathbf{r}) \middle| N \right\rangle, \quad (4)$$

a necessary step is to expand the Fermi field operator $\hat{\Psi}(\mathbf{r})$ appearing in definition (4) on a basis set of single-particle orbitals. A natural choice is the set of Hartee-Fock (or Kohn-Sham) orbitals $\phi_\alpha(\mathbf{r})$:

$$\hat{\Psi}(\mathbf{r}) = \sum_\alpha \hat{c}_\alpha \phi_\alpha(\mathbf{r}), \quad (5)$$

where $\hat{c}_\alpha$ is the operator destroying an electron occupying the $\alpha$ th spin-orbital. We also expand the many-body ground state $\left|N\right\rangle$ on the complete set of Slater determinants $\left|\Phi_i^N\right\rangle$ ($i = 1,2,\ldots$) with coefficients $C_i^N$:

$$\left|N\right\rangle = \sum_i C_i^N \left|\Phi_i^N\right\rangle. \tag{6}$$

In Eq. (6) the $\left|\Phi_i^N\right\rangle$'s are obtained by filling in all possible ways with *N* electrons, consistently with Pauli's exclusion principle, the spin-orbitals $\phi_\alpha(\mathbf{r})$. An expression analogous to (6) holds also for $\left|N-1\right\rangle$, with $\left|N-1\right\rangle = \sum_j C_j^{N-1}\left|\Phi_j^{N-1}\right\rangle$. Here we use the same set of spin-orbitals $\phi_\alpha(\mathbf{r})$ for both *N* and *N – 1* electrons. By inserting (5) and (6) into (4) one obtains an explicit expression for the h-QPWF:

$$\varphi(\mathbf{r}) = \sum_{i,j}(C_j^{N-1})^* C_i^N \sum_\alpha \phi_\alpha(\mathbf{r})\left\langle\Phi_j^{N-1}\middle|\hat{c}_\alpha\middle|\Phi_i^N\right\rangle. \tag{7}$$

Note that the matrix element between Slater determinants occurring in Eq. (7), $\left\langle\Phi_j^{N-1}\middle|\hat{c}_\alpha\middle|\Phi_i^N\right\rangle$, is easily evaluated and takes only the values 0, +1, -1. The result of Eq. (7) is the key for predicting QPWF images.

In the absence of correlation, both ground states $\left|N\right\rangle$ and $\left|N-1\right\rangle$ are each given by a single Slater determinant (assuming that $\left|N\right\rangle$ has a closed shell configuration and neglecting relaxation of Hartree-Fock orbitals between *N* and *N* – 1). Then $C_i^N = \delta_{i,1}$, $C_j^{N-1} = \delta_{j,1}$, and Eq. (7) is reduced to $|\varphi(\mathbf{r})| = |\phi_{\overline{\alpha}}(\mathbf{r})|$, where the index $\overline{\alpha}$ points to the HOMO.

In the generic correlated case, $\varphi(\mathbf{r})$ is a sum over many different orbitals $\phi_\alpha(\mathbf{r})$, with coefficients determined by the many-body expansions of both $\left|N\right\rangle$ and $\left|N-1\right\rangle$. We note incidentally that this is reasonable feature, since STS imaging involves fluctuations of the electron number between *N* and *N* – 1. Furthermore, the absolute value for the coefficient for $\phi_{\overline{\alpha}}(\mathbf{r})$ is generically less than one. Overall, the effect of correlation on the QPWF is twofold:[12]

(i) $\varphi(\mathbf{r})$ is distorted with respect to $\phi_{\overline{\alpha}}(\mathbf{r})$ due to the interference among different orbitals –controlled by correlation- corresponding to the possible transition paths between electron configurations.

(ii) The normalization of $|\varphi(\mathbf{r})|^2$ is smaller than one: the stronger the correlation, the smaller the norm.

In this paper we consider the numerical evaluation of $\varphi(\mathbf{r})$ for representative single molecules adsorbed on a surface.[44] This of course requires to implement a certain set of approximations with respect to the general theory outlined above. In brief:

(1) The molecules are assumed to be well isolated electrically from the substrate on which they are adsorbed. Therefore, the many-body ground states of interest, $|N\rangle$ and $|N-1\rangle$, are those of the isolated interacting molecule. This prohibits the calculation from reproducing the finite width of molecular levels due to tunneling from / to the substrate.

(2) The many-body ground states $|N\rangle$ and $|N-1\rangle$ are obtained from standard ab-initio correlated methods which provide the expansion coefficients $C_i^N$ and $C_j^{N-1}$ entering Eq. (7), such as single and double configuration interaction (CISD) and coupled cluster (CC). This topic is treated in detail in Sec. III. For both methods the single-particle basis set of orbitals $\phi_\alpha(\mathbf{r})$ is truncated with respect to the numerable, complete set. Similarly, the sets of coefficients $C_i^N$ and $C_j^{N-1}$ are subsets of those numerable corresponding to the ideal case of completeness of the Slater-determinant space.

## III. Computational method

### III.1 Calculation of the CI expansion coefficient in Eq. (7)

All the calculations of this paper have been performed with Gaussian 03 on the isolated molecules, i.e., neglecting the effects of the supporting substrate. To check the effects of the substrate (a NaCl layer in the available STM experiments) on our results we have performed calculations on divinylbenzene by including the NaCl layer by a Quantum Mechanical/Molecular Mechanics (QM/MM) method. In practice, the molecule is treated at HF level including in its Hamiltonian the electrostatic interaction with point charges simulating the $Na^+$ and $Cl^-$ ions. In turn, the QM/MM method was preliminary checked by a full QM calculation on divinylbenzene on a $Na_2Cl_2$ cluster. In the QM/MM calculation, a NaCl cluster 4 layer thick, 30 Å x 30 Å in the surface plane simulates the NaCl(100) surface. The distance between the molecular plane and the upper atomic layer of the NaCl surface was 3.0 Å. The resulting HOMO and LUMO (not shown) are indistinguishable from those of the isolated molecule, showing that the substrate effects are inconsequential.

The procedure to calculate CI coefficients is as follow. First, molecular geometries have been optimized at the Hartree-Fock cc-pVDZ level, with the exception of the iodine-containing molecule (for which LANL2DZ has been used) and the nitro-containing molecules, for which 3-21G* provided closer-to-planar geometries than cc-pVDZ (for 4,4 dinitro biphenyl and dinitro-naphtalene planarity problem were not an issue, therefore cc-pVDZ has been used). Then, a closed shell calculation of the *N* state of the molecule at the Hartree-Fock level of theory was performed followed by a CISD calculation for this *N* state.

For the *N* - 1 and *N* + 1 state, CISD calculations were performed on the basis of the HF molecular orbitals obtained for the *N* state of the molecule. To this aim, the local version of Gaussian 03 has been modified to bypass the SCF stage and any modification of the read molecular orbitals. Notably, using the *N*-state molecular orbitals for $N$ - 1 and $N$ + 1 states implies that the orbital relaxation due to the addition of one hole or one electron will be accounted for by the multi-determinantal expansion. Already at this stage, the superiority of (truncated) CC vs (truncated) CI methods is apparent, since relaxation effects are effectively accounted for the single-amplitudes of the CC expansion.

Gaussian provides in output the coefficients of singly and doubly excited configurations both for CISD and CCSD. Due to the computational cost, in the following steps only the configurations with square modulus higher than $10^{-7}$ have been considered. We tested this threshold on a few molecules, and it was found that configurations with square modulus lower than $10^{-7}$ did not cause dramatic changes of the results. For CCSD calculations, we remark that we did not use the entire CCSD wavefunction, but considered a multi-configuration expansion including only single and double excitations.

For the CI and CC calculations we have used the $10^{-7}$ convergence criteria for iterative Davidson diagonalization in Gaussian. For the simplified model of Ni ($L_2$) (see Figures 9 and 10), convergence was achieved only for a $10^{-6}$ threshold.

Correlated methods require large basis sets.[47] We have tested several basis sets, and we found that the cc-pVDZ represents the best compromise between accuracy and computational costs. A comparison between cc-pVDZ and cc-pVTZ results is described in Sect. IV as well the in the Discussion S2 of the Supplemental Material.

The CI coefficients provided in the Gaussian output are read by a parallel home-made Fortran code that yields the QPWF in Eq. (7) spanned over the *N*-state molecular orbitals,[48] and also over the original atomic orbital basis sets, to be manipulated as described below to provide the images reported in Sect. IV.

### III.2 Visualization of the QPWFs

The QPWF spanned over the atomic basis set is inserted into a check-point Gaussian file. The latter can be manipulated with the cubegen Gaussian utility to create a cube file of the QPWF that can be visualized with many graphic tools, including VMD. Further manipulations of the QPWFs are possible with cubegen, such as to take the square modulus of the QPWF to provide the spectral density, and thus constant-height

STM images.[1] Since QPWFs contain more information than spectral densities, in this paper, where a direct comparison with experiments is not discussed, we shall present only QPWFs (and not spectral density) results.

As it is described in Sec. IV, it was found that the QPWFs resulted to be very similar to the corresponding Hartree-Fock HOMO and LUMO. Therefore, to visualize the effect of correlation additional images were produced by taking the difference between HOMO and h-QPWF (a quantity called SUBH) and between LUMO and e-QPWF (SUBL) with the Gaussian cubman utility. This difference is the more straightforward way to analyze the relation between QPWFs and frontier orbitals. However, we found that SUBH and SUBL are generally dominated by the corresponding frontier orbitals and they appear as scaled version of HOMO and LUMO. To highlight the contribution of all the orbitals except HOMO and LUMO, we defined two other quantities, called h-QPWF-ZERO and e-QPWF-ZERO, defined to be the projections of h-QPWF and e-QPWF in the subspace orthogonal to HOMO and LUMO, respectively.

All these three-dimensional quantities (HOMO, LUMO, QPWFs, SUBH/L, QPWFs-ZERO) were imaged as color maps on a planar slice. This slice was taken 3.0 Å over the molecular plane, parallel to it, for the frontier orbitals (i.e., HOMO and LUMO) and the QPWFs, while it was taken at 1.0 Å for SUBL, SUBH, and QPWF-ZERO. The color scale maximum (red) and minimum (blue) were $10^{-2}$ Å$^{-3/2}$ and $-10^{-2}$ Å$^{-3/2}$ for the HOMO, LUMO, the QPWFs, SUBH, and SUBL. On the other hand, the h-QPWF-ZERO and e-QPWF-ZERO images in all cases were obtained setting the color scale extrema at $10^{-5}$ Å$^{-3/2}$ (see Figure 1), unless otherwise noted.

## IV. Results and Discussion

### IV.1 Planar conjugated molecules

Aromatic molecules, such as benzene derivatives, represent simple systems to test the developed method. In fact, they are often planar, that simplifies the analysis of the resulting QPWFs, and they can be substituted by several different chemical groups, that allows verifying how the substituent inductive and mesomeric effects modify correlation. We first consider para-divinylbenzene, a benzene ring with two vinyl substituents in opposite positions. We arrange substitutions in a $C_{2h}$ symmetry. Figure 1 shows the contour plots of selected slices of the HOMO and LUMO at the Hartree-Fock level (first column of Fig. 1) as well as their corresponding h-QPWF and e-QPWF (second column of Fig. 1) calculated at the CISD level. The color code in Fig. 1 is such that the red (blue) color stands for positive (negative) values of the wave function, whereas white regions locate nodal surfaces.

As explained in Sec. II, in a single-particle (i.e., uncorrelated) picture the h-QPWF reduces to the HOMO and the e-QPWF to the LUMO. Therefore, the differences between HOMO and h-QPWF (between LUMO and e-QPWF) point to contributions beyond Hartrtee-Fock theory, that we ascribe to electron correlation. Since there are no visible differences between the Hartree-Fock orbitals and QPWFs of Fig. 1 we conclude that correlation effects have a minor impact on this molecule and may not be resolved by current experimental orbital imaging techniques. Nevertheless, it is conceptually relevant to single out the signatures of correlation emerging in wave function images, as we discuss below.

To highlight the discrepancies between QPWF and Hartree-Fock orbitals, we plot in the last column of Fig. 1 the image of the difference between the HOMO (LUMO) at the Hartree-Fock level and the h-QPWF (e-QPWF) [labeled in Fig. 1 as SUBH (SUBL)]. Since the QPWFs appear to just be scaled versions of HOMO and LUMO, it is not surprising that SUBH and SUBL are also similar to HOMO and LUMO. To investigate the contributions to QPWF brought from orbitals other than the frontier ones, the third column of Fig. 1 shows two functions obtained from the QPWFs by omitting the dominant contributions of the HOMO and LUMO (h-QPWF-ZERO and e-QPWF-ZERO, respectively, cf. Sec. III). These QPWF-ZERO contour plots directly image in real space the distortion of QPWFs with respect to frontier orbitals, an effect of electron correlation beyond the simple HOMO and LUMO re-scaling. Noticeably, the QPWF-ZERO wave function structure may be resolved only reducing the color scale by three orders of magnitude with respect to the first and second columns of Fig. 1 (cf. also Table I). This illustrates quantitatively that the discrepancies between HOMO and h-QPWF (between LUMO and e-QPWF) are minimal.

In the h-QPWF-ZERO image of Fig. 1 we observe that the QPWF weight is substantial over the $\pi$-cloud of the benzene ring and also in the two vinyl groups of the system. In the e-QPWF-ZERO image orbital lobes originating from correlation extend mainly across the benzene carbon atoms. In order to decrease the system symmetry and hence investigate whether this affects the discrepancies occurring between Hartee-Fock frontier orbitals and QPWFs, we apply our method to a significant set of divinylbenzene derivatives. Specifically, we replace the hydrogen atom of one of the vinyl groups with an element of the halogen series F, Cl, Br, I, as well as with the cyano (-CN), thiol (-SH), hydroxyl (-OH) and nitro (-$NO_2$) groups. The rationale is to study the evolution of QPWFs as we include substituents of different electronegativity that may differently modulate electron correlation effects. The images obtained for the halogen series F, Cl, Br, I (cf. Figures S1, S2, S3, S4, respectively, in the Supplemental Material), as well as with the cyano (-CN) (Fig. S5), thiol (-SH) (Fig. S6), and hydroxyl (-OH) (Fig. S7), showed similar behavior to those predicted for para-dininylbenzene (cf. Fig. 1), although some minor discrepancies have been observed (see also Discussion S2 in the Supplemental Material).

The case of the nitro-divinyl benzene derivative is interesting since the h-QPWF-ZERO and e-QPWF-ZERO images (presented in Fig. 2) behave differently from the corresponding images of the other divinylbenzene derivatives. In fact, with respect to the bromine (see Fig. S3) and cyanide (see Fig. S5) derivatives the orbital lobes of h-QPWF-ZERO on the position 2 and 6 of the benzene ring are larger whereas the corresponding orbital lobes in the position 3, 4 and 5 are less dense (cf. labels in Fig. 2). This behavior is due to the presence of the nitro group which is an electron acceptor. Additionally, from the e-QPWF-ZERO plot of Fig. 2 we observe that the orbital lobes behave differently in comparison to the other derivatives. This difference is mainly caused from the contribution of the LUMO+1 (cf. Table I). This has been verified by creating an image by setting zero not only the contribution of the LUMO but also the contribution of the LUMO+1, which results in orbital lobes similar to those obtained for the other derivatives. Note that nitro-divinylbenzene is not planar as there is a distortion of the angle that defines the free vinyl group around $20^{\circ}$ degrees (C1-C7-C8-H9).

In order to substantiate quantitatively our findings on QPWFs images, in Table I we report the computed expansion coefficients of QPWFs referred to the molecular-orbital basis set [made of the $\phi_\alpha(\mathbf{r})$'s occurring in Eq. (7)] for the 7 of the 18 molecules examined in this paper (the expansion coefficients of QPWFs of the rest of the molecules examined are presented in table SI in the Supplemental Material). Since the basis-set size may be very large, we consider only the most important orbital coefficients chosen in the range of the lowest (highest) eleven orbitals starting from the LUMO (ending with the HOMO). The analysis of these coefficients confirms that the modifications of QPWFs of divinylbenzene and their derivatives due to electron correlation are mainly a renormalization of HOMO and LUMO, while the contribution from other orbitals is minimal. In fact, the main contributions to QPWFs regard the Hartree-Fock HOMO and LUMO, their coefficients being typically three orders of magnitude larger than those for other orbitals (cf. Table I).

With respect to the molecules tested so far, the effect of correlation is stronger in the nitro derivative of divinylbenzene, for which the $NO_2$ group decreases the coefficients of HOMO and LUMO (around 0.85) in comparison to the other divinyl-benzene derivatives (see Table SI in the Supplemental Material). For this reason we additionally investigate nitrobenzene as well as the systematic addition of nitro groups to the benzene molecule (i.e. di-nitrobenzene and tri-nitrobenzene molecule). Figure 3 presents the orbital images for nitrobenzene. The CISD calculations were performed using the cc-PVDZ basis set. Since the frontier orbitals at the Hartree-Fock level and the QPWFs are almost indistinguishable, we focus on the h-QPWF-ZERO and e-QPWF-ZERO images, where dense orbital lobes are observed between the positions 2-3 and 5-6 of the phenyl ring. This explains the slight shrinkage of the h-QPWF and e-QPWF images. The CISD calculations were also repeated by employing the larger basis set cc-PVTZ (see Figure S8 in the Supplemental Material). The results obtained were found very similar to that obtained using the cc-PVDZ basis set although the orbital lobes of the h-QPWF and the e-QPWF are additionally reduced. The images of the HOMO and LUMO and the corresponding QPWFs of dinitrobenzene and trinitrobenzene as well as of a larger molecule containing two phenyl and two nitro groups (4,4 dinitro biphenyl) were also calculated (see Figures S9, S10, and S11, respectively, in the Supplemental Material). The scenario illustrated in Figure 3 is confirmed in all these cases although a small discrepancy is depicted between the LUMO and the e-QPWF of (4,4 dinitro biphenyl) (cf. Fig S11 in the Supplemental Material).

## IV.2. Poly-aromatic molecules

In this section we consider poly-aromatic molecules, starting from the simplest case of naphthalene, which is characterized by two fused benzene rings. As for molecules of Sec. IV.1, also QPWFs of naphthalene are very similar to the HF frontier orbitals (See Figure S12 in the Supplemental Material). To test the effect of decreasing the molecular symmetry we also replace two hydrogen atoms with the nitro group. Figure 4 represents the HOMO and the LUMO at the HF level and the CISD QPWFs for this derivative. By directly comparing the HF orbitals and the QPWFs we notice the shrinkage of the orbital lobes of the QPWFs -especially the e-QPWF. Additionally, from QPWF-ZERO images we observe the effective reduction of symmetry due to the two nitro groups.

An important representative of the class of poly-aromatic molecules is pentacene, since this has been studied both experimentally and theoretically.[20] Specifically, experimental images substantially agree with Hartree-Fock / density-functional-theory results, demonstrating that pentacene does not present important correlation effects. Therefore, we expect that the correlated QPWFs are not qualitatively different from Hartree-Fock HOMO and LUMO orbitals. This is illustrated in Fig. 5, where both the HOMO and the LUMO are essentially the same as the QPWFs, except for a small shrinkage of the orbital lobes. Such behavior is similar to that of napthalene and dinitro-naphathalene discussed above. We also observe that the QPWF of pentacene must satisfy a large number of symmetry constraints. This may further suppress correlation effects connected to changes of the shape and topology of the orbital lobes.

## IV.3. A planar metal complex

In this section we investigate one example of planar molecule that contains a metal center. Even if the computational effort to study such a molecule is bigger than for the simpler complexes studied above, due to the presence of the metal, nevertheless this case study is relevant since the valence electrons of the metal are expected to experience significant correlation. Specifically, we focus on the planar nickel complex $Ni(L)_2$ (L=3,5-di-tert-butyl-o-diiminobenzosemiquinonate(1-)) (see Reference 50 for a previous study). Ni complexes, although of other type, have been recently imaged by STM on insulating layers.[21] From Fig. 6 one may see that a few small differences between the Hartree-Fock LUMO and e-QPWF plots emerge, in particular some loss of weight is clearly visible in the e-QPWF. The QPWF-ZERO images show dense orbital lobes that arise from the metal center in both the QPWFs. Besides, the SUBH and SUBL images illustrate that the modifications of the Hartree-Fock orbital lobes mainly consist of rescaling of the molecular orbitals that define the HOMO and LUMO, with the exception that in the SUBL image additional weight is observed around the Nickel atom.

Additionally, from Table I we observe that for this molecule the contribution of the HOMO and LUMO Hartree-Fock orbitals to QPWFs is in both cases around 0.77, which is less than the corresponding coefficients found for the other molecules examined. We believe these results highlight the role of complexes that contain a metal atom as promising candidates to show correlation effects in STM imaging.

### IV.4. Coupled-cluster calculations

The results presented so far were obtained at the CISD level. Due to its high computational cost, it is not always feasible to go beyond this level for the systems studied. Thus, in order to test whether the CISD level is appropriate to compute the QPWFs, we have selected a subset of simple molecules where the calculation performed at the coupled cluster with single and double excitations (CCSD) level may be afforded. Therefore, we reconsider divinylbenzene and the $Ni(L_2)$ complex, first studied at the CISD level in Fig. 1 and Fig. 6 respectively. Figure 7 represents the frontier orbitals of the $Ni(L_2)$ molecule whereas Fig. 8 the frontiers orbitals of divinylbenzene based on the Hartree-Fock (first column) and CCSD (second column) calculations. We find discrepancies between the images obtained for divinylbenzene and $Ni(L_2)$ from the CCSD calculations and those obtained from the CISD calculations. Specifically, the h-QPWF-ZERO image (third column of Fig. 7 and Fig. 8) obtained from the CCSD calculation is qualitatively different from that calculated at the CISD level (third column of Fig. 6 and Fig. 1), showing different nodal contour lines. Additionally, the QPWF-ZERO images obtained from the CCSD calculations are displayed setting the color scale cutoff at $10^{-4}$ Å$^{-3/2}$ in Fig. 7 and Fig. 8, in contrast to the CISD results of Fig. 6 and Fig. 1 where the color scale cutoff was set at $10^{-5}$ Å$^{-3/2}$. This shows that the variation of the electronic probability amplitude of QPWFs due to correlation effects, according to the CCSD calculation, is one order of magnitude larger than the CISD value.

As a further test we consider a molecule derived from the $Ni(L_2)$ complex. Figures 9 and 10 show the frontiers orbitals of the molecule respectively at the CISD and CCSD levels. As for divinylbenzene, the color scale maximum ($10^{-4}$ Å$^{-3/2}$) for the QPWF-ZERO images obtained from the CI calculations shown in Fig. 9 is one order of magnitude smaller than that used for CCSD images ($10^{-3}$ Å$^{-3/2}$) of Fig. 10. Comparing Figs. 9 and 10 we see that more dense orbital lobes can be observed around the Nickel atom and the orbital lobes that define the nitrogen atoms. Additionally, in Table I we observe that the Hartree-Fock coefficients contributing to the orbital expansion of the QPWF are substantially reduced when going at the CC level [from 0.95 at the CI level to 0.80 (h-QPWF) and from 0.94 to 0.77 (e-QPWF)]. On the other hand, for the entire Ni($L_2$) complex the frontier orbital coefficients have been reduced only slightly in going from CI to CC calculations [from 0.774 at the CI level to 0.770 (h-QPWF) and from 0.77 to 0.76 (e-QPWF)]. Although it is not easy to identify the reason for this difference [the simplified Ni($L_2$) molecule has intrinsically different orbitals than Ni($L_2$)], it seems that the higher delocalization allowed by the aromatic rings in the original Ni($L_2$) induces an orbital energy ladder with a finer spacing that favor correlation effects.

## V. Conclusion

In this work we have presented a many-body theory of tunneling developed to show how electronic correlation can be visualized in STS images of molecules decoupled from the supporting conductive substrate. Effects of electronic correlation have been already demonstrated to be important for the molecular energy levels. Here, we have focused on the distortion of STS images operated by electronic correlation, investigating the differences between many-body STS-related quantities (QPWFs) and their uncorrelated counterparts (HF frontier orbitals). Starting from the theoretical equations, we have set up a protocol that yields QPWFs and spectral densities on the basis of standard Quantum Chemical calculations performed with the widespread Gaussian code. These QPWFs can be easily visualized by common molecular graphics codes such as VMD. Due to the computational costs, the Quantum Chemical method considered in this paper for most of the calculations has been at the minimal level needed to grasp correlation effects (CISD). However, the possibility of using higher level calculations (CCSD) has been also explored. This has shown that to obtain realistic results it is necessary to go beyond the minimal CISD level.

For the investigated molecules, we have found, not unexpectedly, that the QPWFs basically correspond to shrunk HF frontier orbitals, although other orbitals can contribute (up to a few percents for the studied systems). These contributions have been highlighted by defining a QPWF-derived quantity (QPWF-ZERO) that allows for a molecular level analysis of electronic correlation effects. For the studied molecules, we found distortions of the STS images that are too small to be verifiable by current experimental setups. Nevertheless, our calculations provided some hints to go in the direction of larger correlation effects: it is clear that systems of high symmetry should be avoided as the symmetry constrains reduce the allowed distortions. Moreover, metal complexes appear quite promising, since the highest

QPWF-HF orbital differences were found for $NiL_2$. Following these guidelines and exploiting the developed theoretical and computational tools, our future research will consider other molecules to seek out systems with experimentally detectable correlation effects.

## Acknowledgements

The authors thank Germar Hoffmann, Elisa Molinari, and Roland Wiesendanger for useful discussions. Funding from INFM under the Young Researcher Seed Project 2008 initiative is gratefully acknowledged. Computer time has been provided by CINECA (http://www.cineca.it) under the 2008-2009 INFM supercomputer project grants.

## Figures

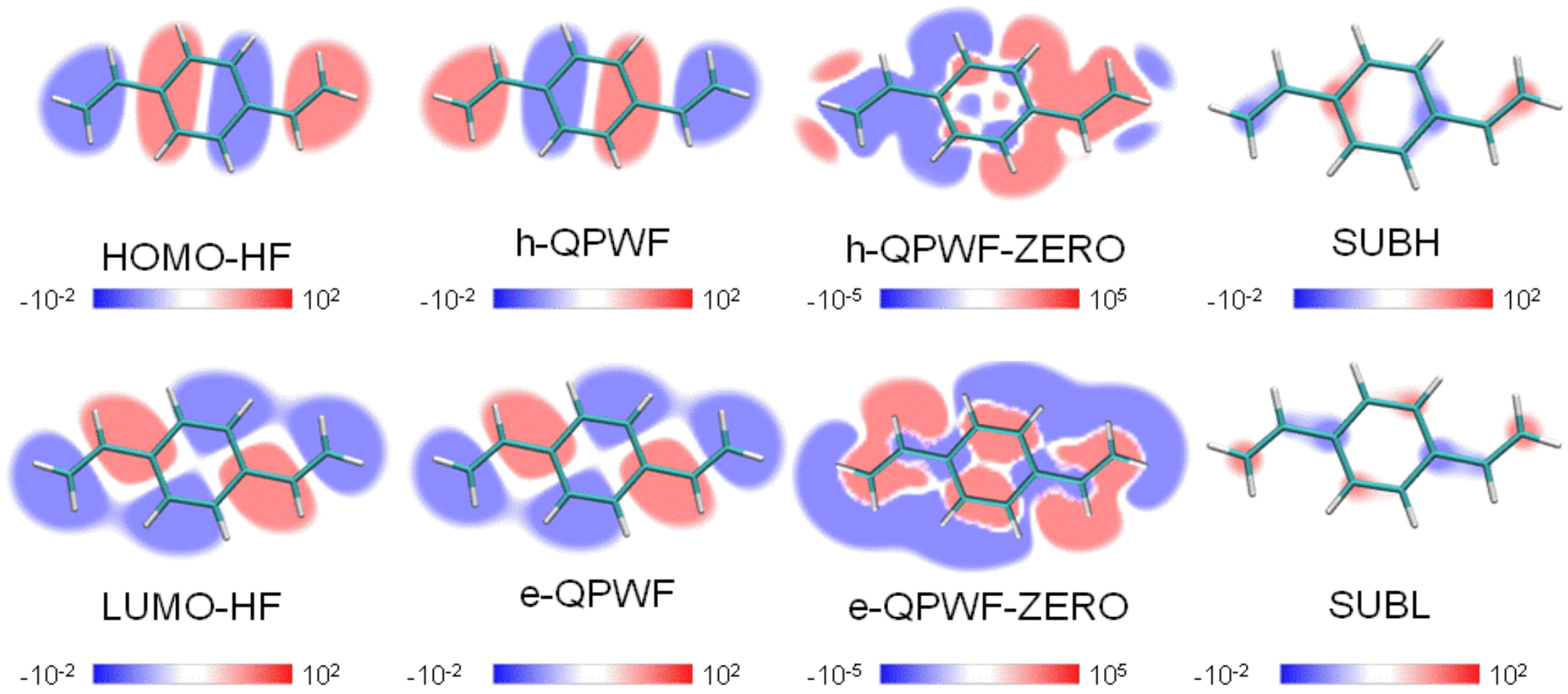

**Fig 1.** Slices of frontier orbitals (i.e., HOMO and LUMO), QPWFs (h-QPWF and e-QPWF), QPWFs omitting the contribution on the HOMO and LUMO orbital (h-QPWF-ZERO and e-QPWF-ZERO), and result of the subtraction between the HOMO and h-QPWF-ZERO (SUBH) and the LUMO with e-QPWF-ZERO (SUBL) of divinylbenzene.

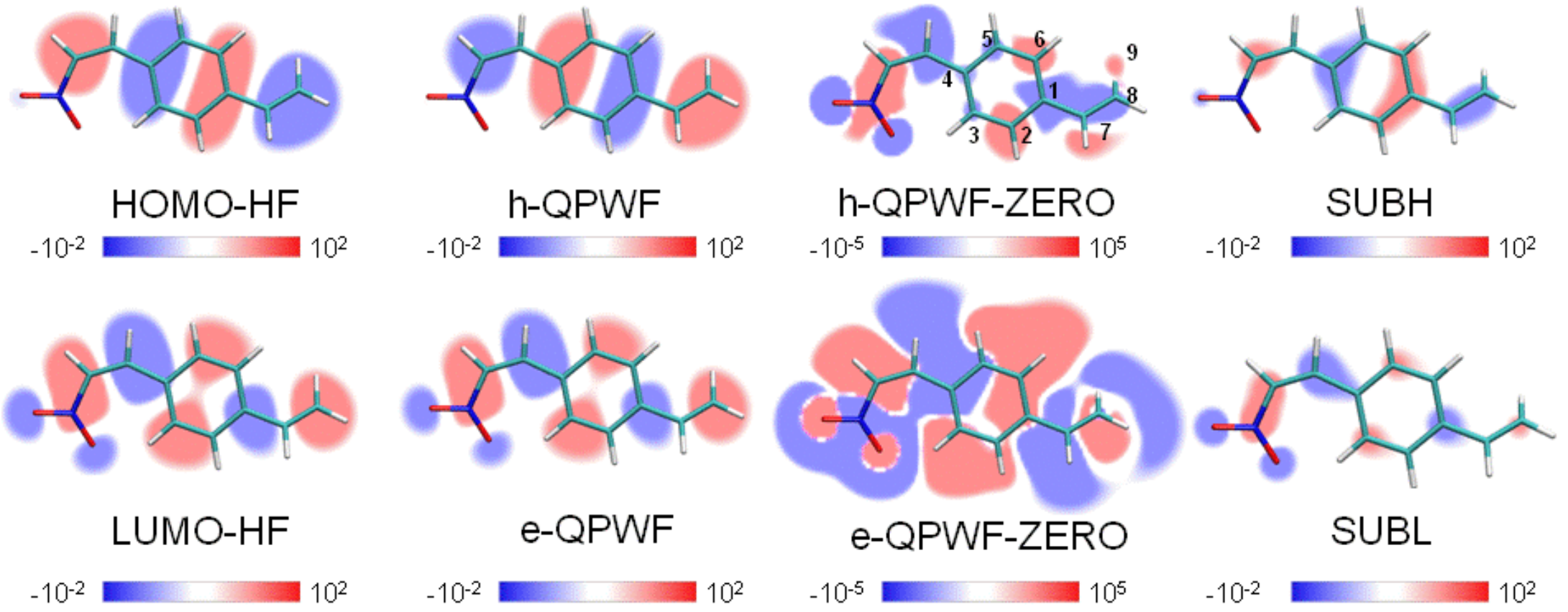

**Fig 2.** Slices of frontier orbitals (i.e., HOMO and LUMO), QPWFs (h-QPWF and e-QPWF), QPWFs omitting the contribution on the HOMO and LUMO orbital (h-QPWF-ZERO and e-QPWF-ZERO), and result of the subtraction between the HOMO and h-QPWF-ZERO (SUBH) and the LUMO with e-QPWF-ZERO (SUBL) of nitro-divinylbenzene.

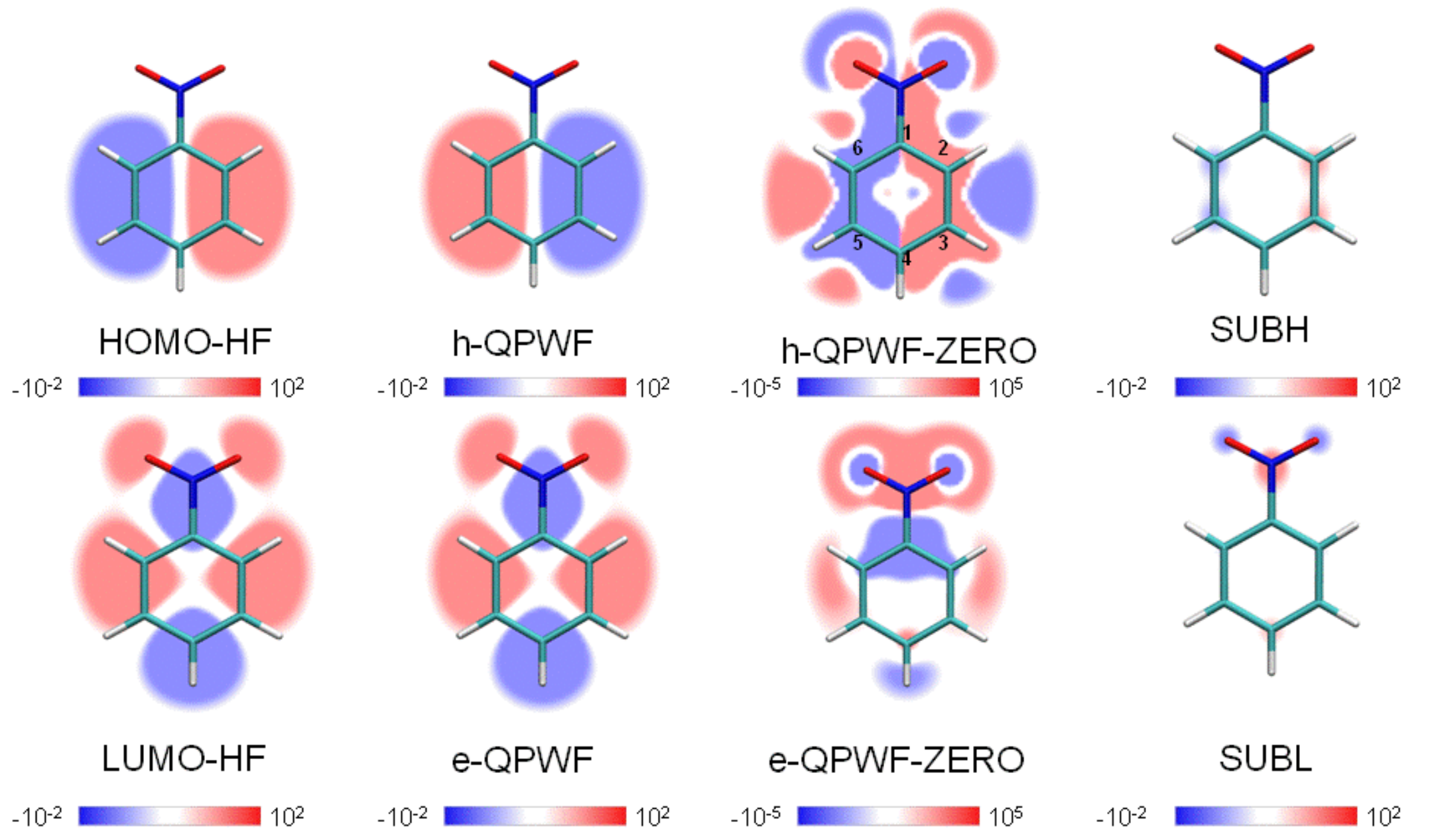

**Fig 3.** Slices of degenerate frontier orbitals (i.e., HOMO and LUMO), QPWFs (h-QPWF and e-QPWF), QPWFs omitting the contribution on the HOMO and LUMO orbital (h-QPWF-ZERO and e-QPWF-ZERO), and result of the subtraction between the HOMO and h-QPWF-ZERO (SUBH) and the LUMO with e-QPWF-ZERO (SUBL) of nitrobenzene.

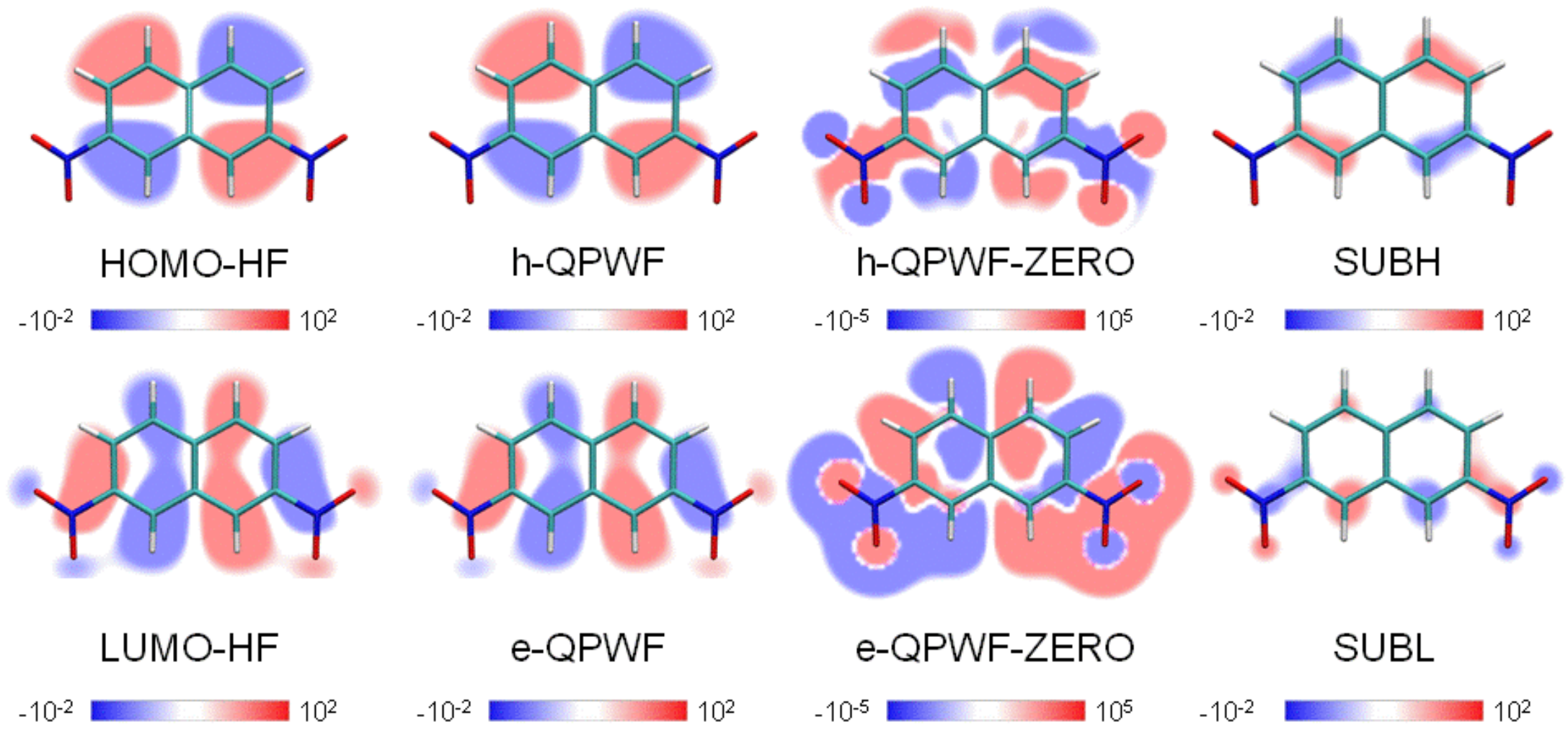

**Fig 4.** Slices of frontier orbitals (i.e., HOMO and LUMO), QPWFs (h-QPWF and e-QPWF), QPWFs omitting the contribution on the HOMO and LUMO orbital (h-QPWF-ZERO and e-QPWF-ZERO), and result of the subtraction between the HOMO and h-QPWF-ZERO (SUBH) and the LUMO with e-QPWF-ZERO (SUBL) of 2-7-dinitro-naphthalene.

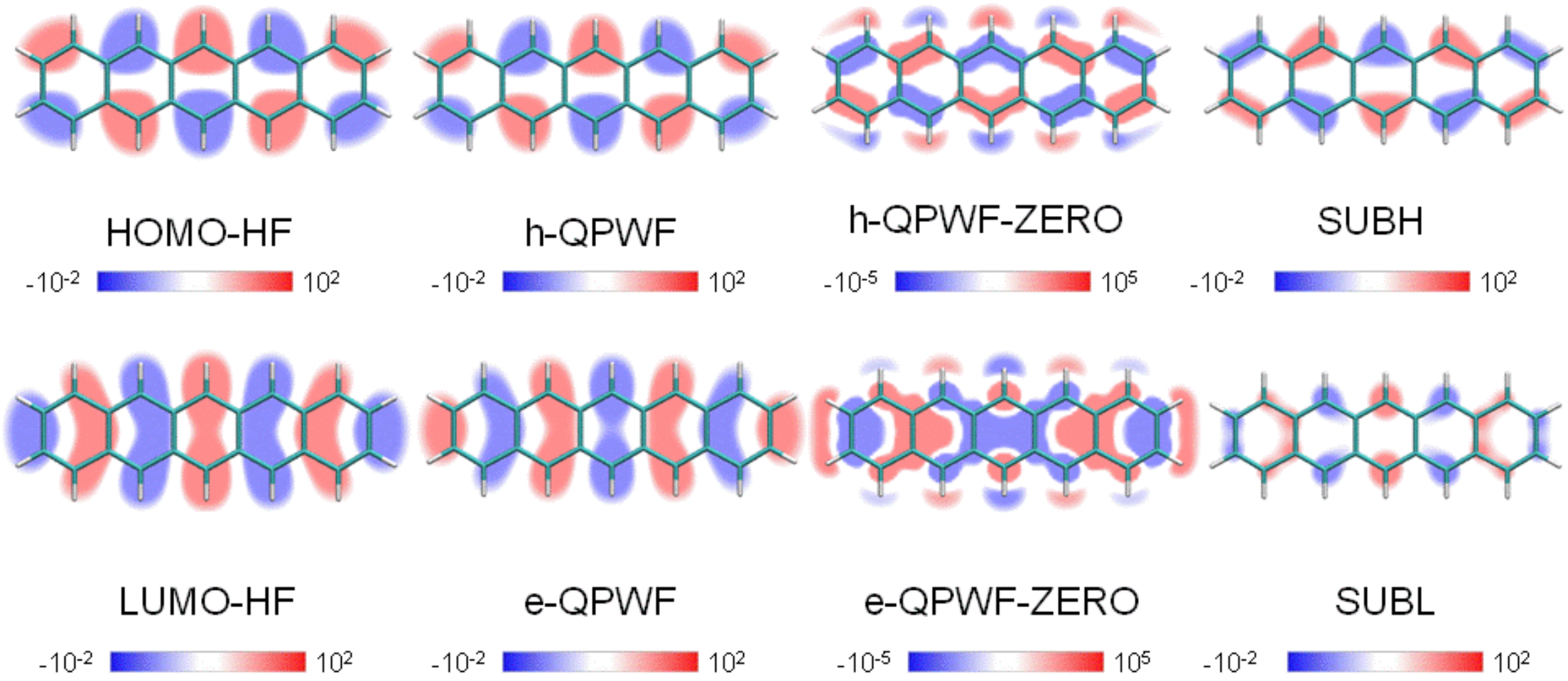

**Fig 5.** Slices of frontier orbitals (i.e., HOMO and LUMO), QPWFs (h-QPWF and e-QPWF), QPWFs omitting the contribution on the HOMO and LUMO orbital (h-QPWF-ZERO and e-QPWF-ZERO), and result of the subtraction between the HOMO and h-QPWF-ZERO (SUBH) and the LUMO with e-QPWF-ZERO (SUBL) of pentacene.

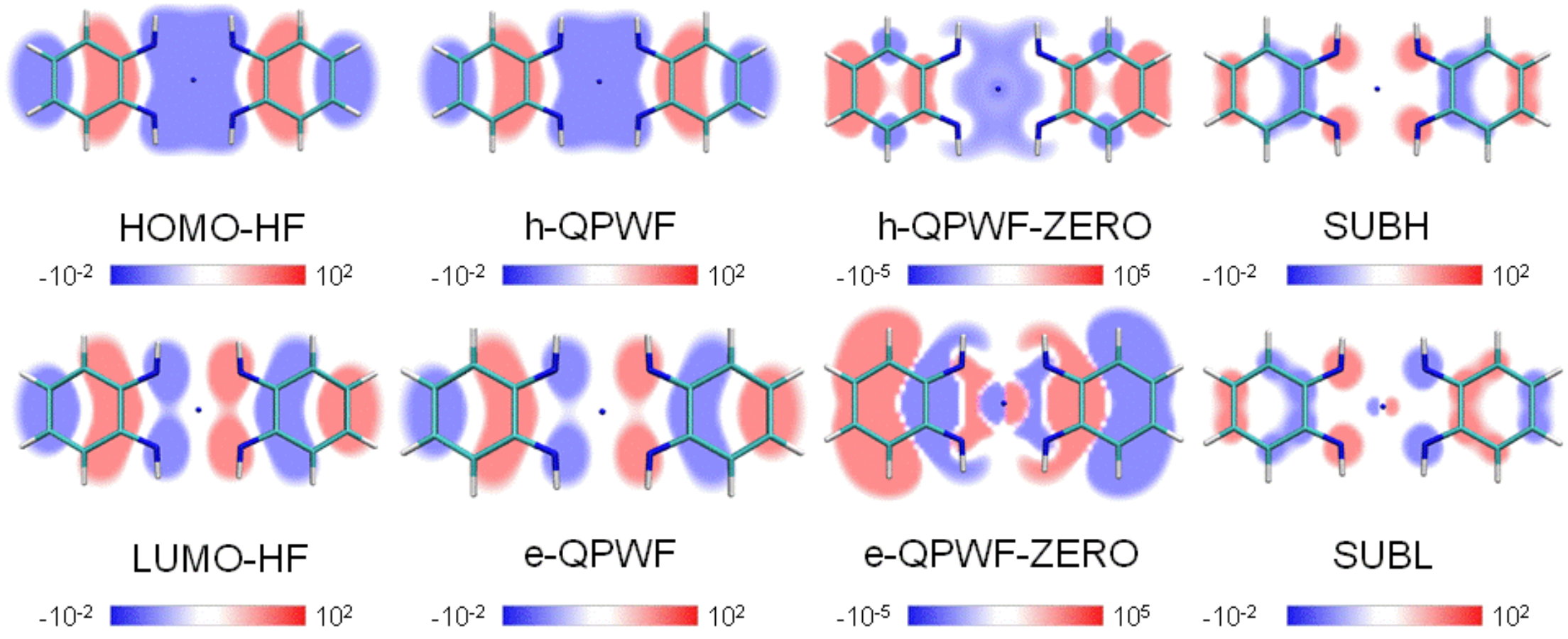

**Fig 6.** Slices of frontier orbitals (i.e., HOMO and LUMO), QPWFs (h-QPWF and e-QPWF), QPWFs omitting the contribution on the HOMO and LUMO orbital (h-QPWF-ZERO and e-QPWF-ZERO), and result of the subtraction between the HOMO and h-QPWF-ZERO (SUBH) and the LUMO with e-QPWF-ZERO (SUBL) of Ni($L_2$) (L=3,5-di-*tert*-butyl-*o*-diiminobenzosemiquinonate(1-)).

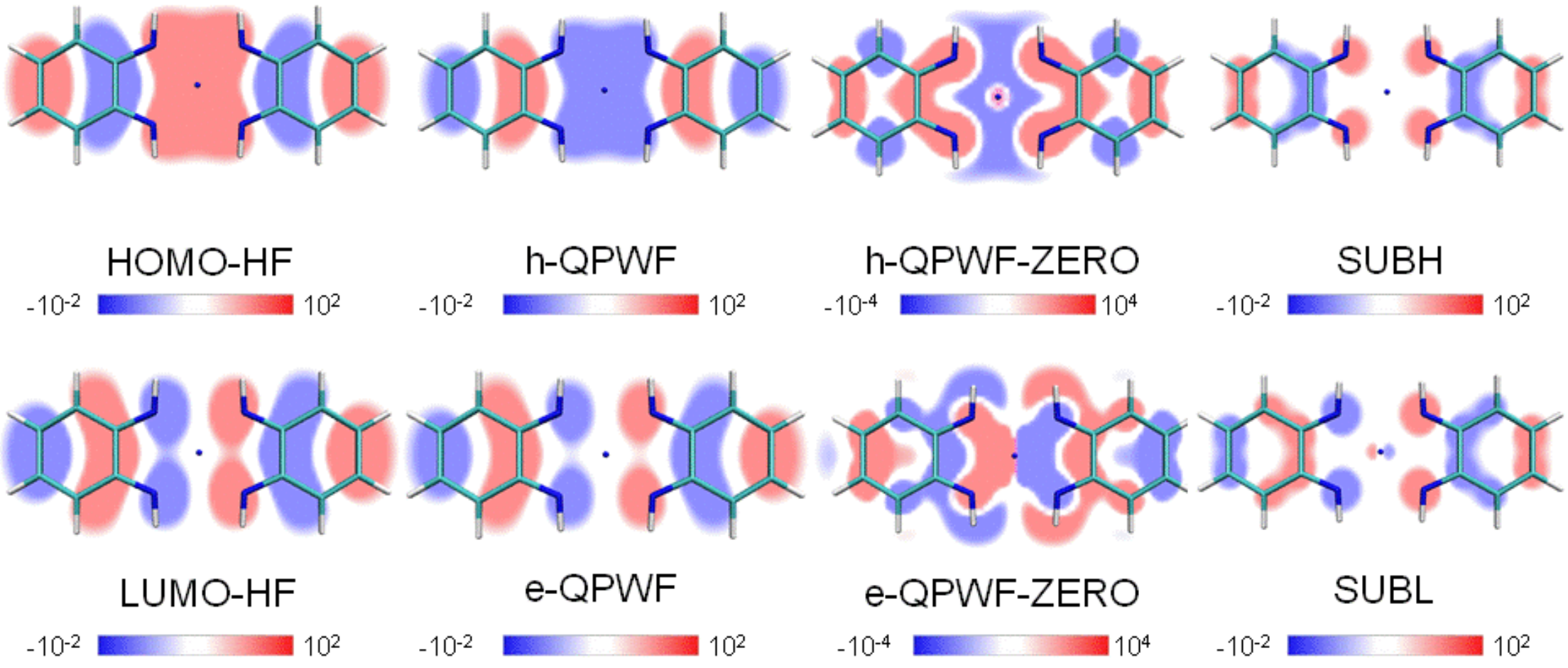

**Fig 7.** Slices of frontier orbitals (i.e., HOMO and LUMO), QPWFs (h-QPWF and e-QPWF), QPWFs omitting the contribution on the HOMO and LUMO orbital (h-QPWF-ZERO and e-QPWF-ZERO), and result of the subtraction between the HOMO and h-QPWF-ZERO (SUBH) and the LUMO with e-QPWF-ZERO (SUBL) of Ni($L_2$) (L=3,5-di-*tert*-butyl-*o*-diiminobenzosemiquinonate(1-)) at the CCSD level of theory.

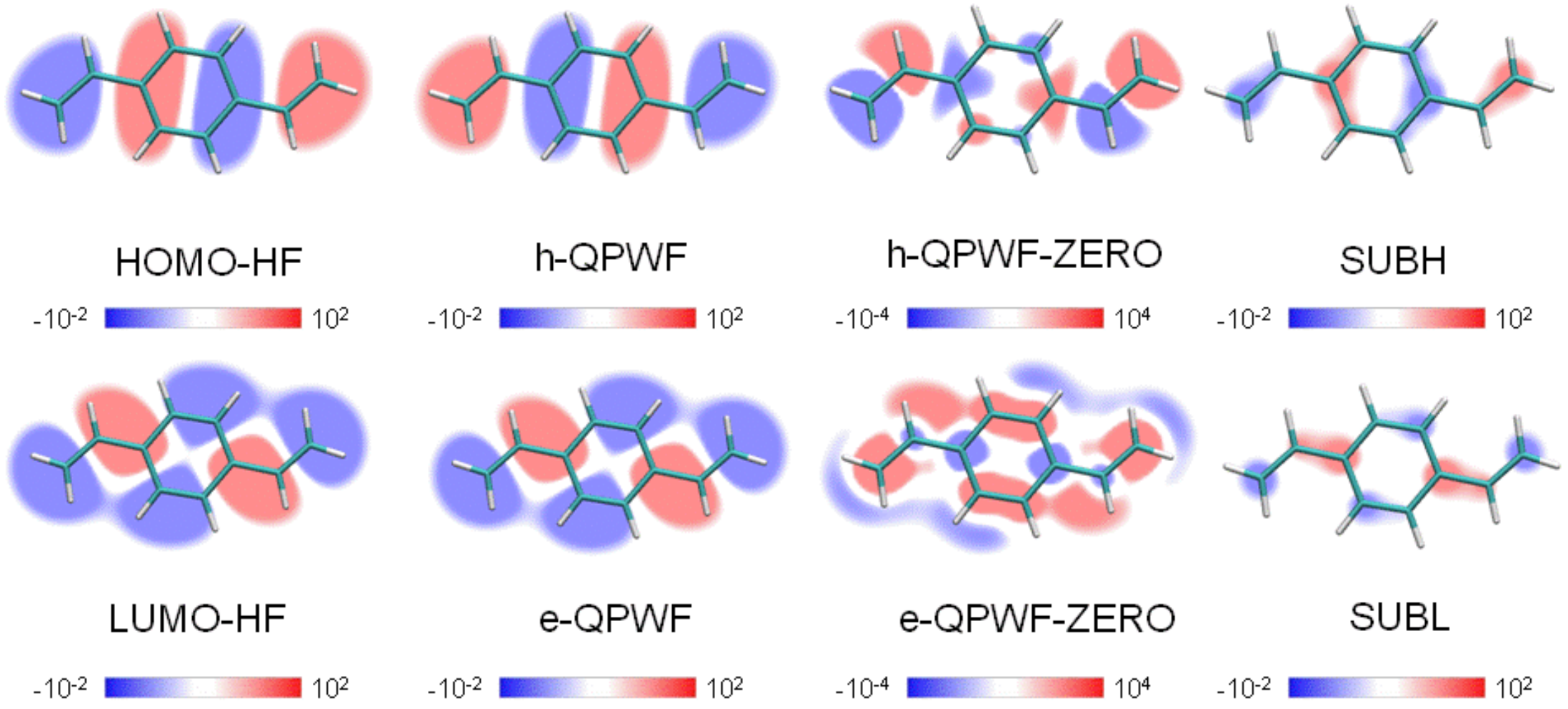

**Fig 8.** Slices of frontier orbitals (i.e., HOMO and LUMO), QPWFs (h-QPWF and e-QPWF), QPWFs omitting the contribution on the HOMO and LUMO orbital (h-QPWF-ZERO and e-QPWF-ZERO), and result of the subtraction between the HOMO and h-QPWF-ZERO (SUBH) and the LUMO with e-QPWF-ZERO (SUBL) of divinylbenzene at the CCSD level of theory.

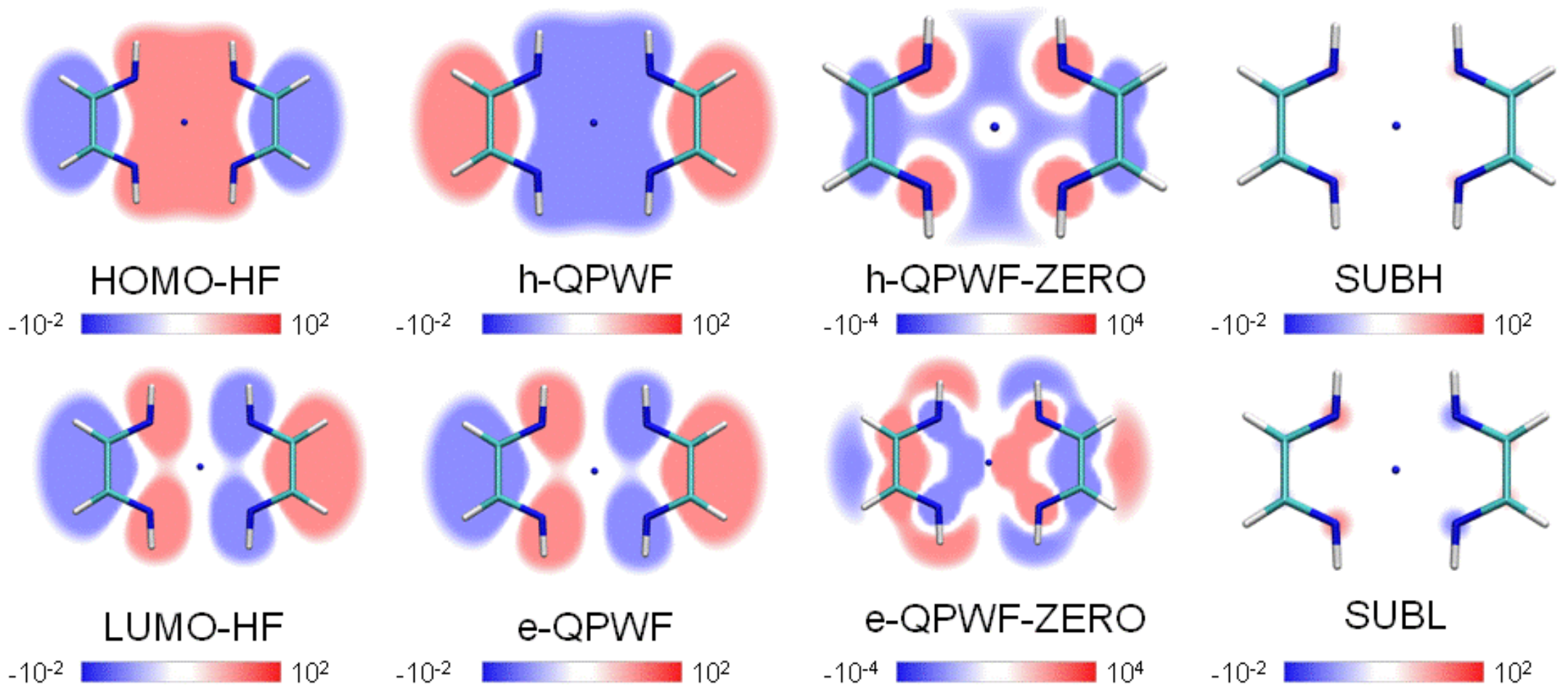

**Fig 9.** Slices of frontier orbitals (i.e., HOMO and LUMO), QPWFs (h-QPWF and e-QPWF), QPWFs omitting the contribution on the HOMO and LUMO orbital (h-QPWF-ZERO and e-QPWF-ZERO), and result of the subtraction between the HOMO and h-QPWF-ZERO (SUBH) and the LUMO with e-QPWF-ZERO (SUBL) of simplified molecule of $Ni(L_2)$ at the CISD level of theory.

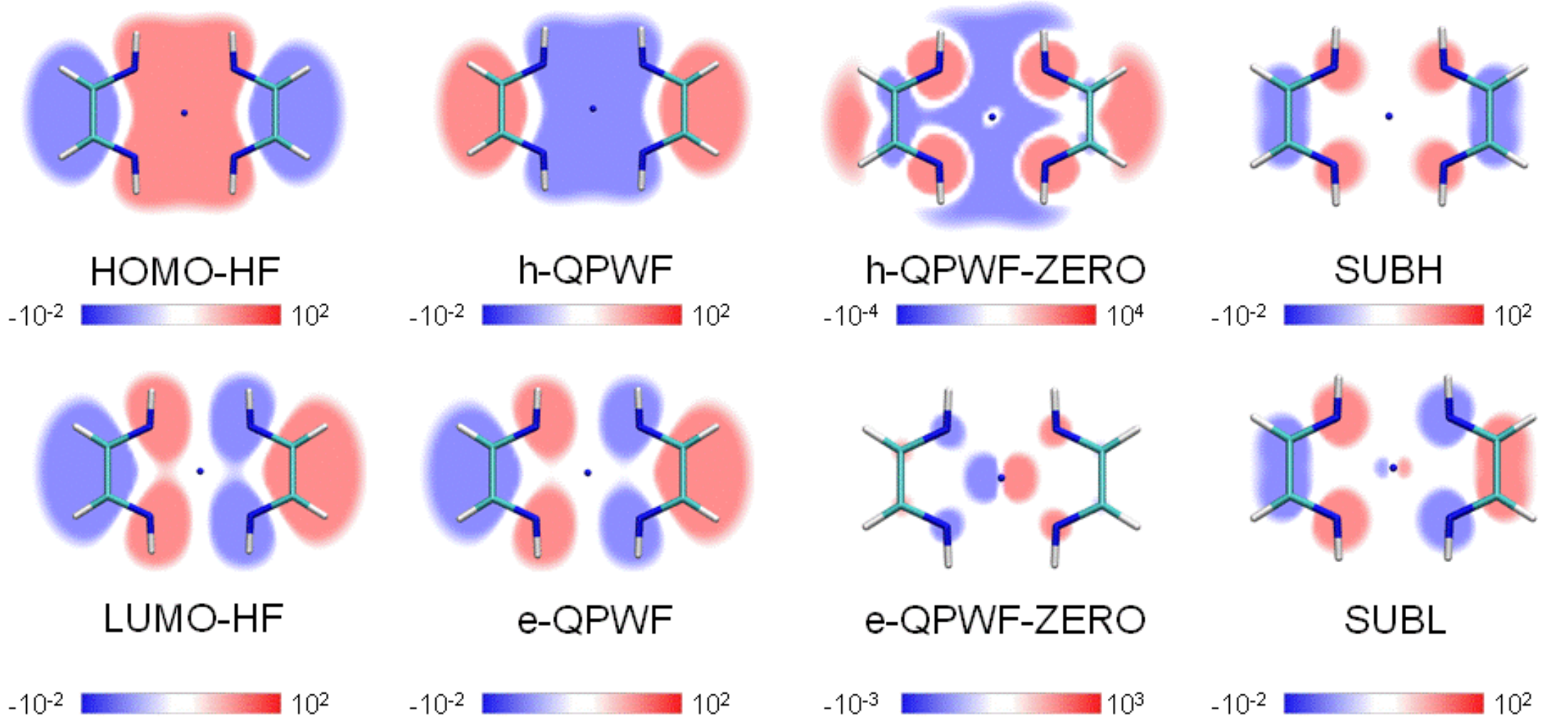

**Fig 10.** Slices of frontier orbitals (i.e., HOMO and LUMO), QPWFs (h-QPWF and e-QPWF), QPWFs omitting the contribution on the HOMO and LUMO orbital (h-QPWF-ZERO and e-QPWF-ZERO), and result of the subtraction between the HOMO and h-QPWF-ZERO (SUBH) and the LUMO with e-QPWF-ZERO (SUBL) of simplified molecule of $Ni(L_2)$ at the CCSD level of theory.

| Molecule | Figure | h-QPWF | e-QPWF |
|---|---|---|---|
| Divinyl-Benzene | 1 | 0.899**(HOMO)**+9.79*$10^{-4}$**(HOMO-3)**+2.99*$10^{-4}$**(LUMO+2)** | 0.895**(LUMO)**+5.22*$10^{-4}$**(HOMO-2**)+2.52*$10^{-4}$**(LUMO+1)** |
| Nitro-Divinyl-Benzene | 2 | 0.859(**HOMO)**+1.17*$10^{-4}$**(HOMO-2)**+8.47*$10^{-4}$**(LUMO)** | 0.854**(LUMO)**+4.29*$10^{-4}$**(HOMO)**+1.92*$10^{-3}$**(LUMO+1)** |
| Nitro-Benzene | 3 | 0.946**(HOMO)**+3.01*$10^{-4}$**(HOMO-2)**+2.29*$10^{-4}$**(LUMO+1)** | 0.940**(LUMO)**+1.01*$10^{-3}$**(HOMO-1)**+2.76*$10^{-3}$**(LUMO+2)** |
| 2-7-Dinitro-Naphthalene | 4 | 0.873**(HOMO)**+5.09*$10^{-4}$**(LUMO)**+4.75*$10^{-4}$**(LUMO+2)** | 0.870**(LUMO)**+1.89*$10^{-4}$**(HOMO)**+1.64*$10^{-3}$**(LUMO+2)** |
| Pentacene | 5 | 0.819**(HOMO)** | 0.817**(LUMO)**+3.53*$10^{-4}$**(HOMO-4)**+1.49*$10^{-4}$**(LUMO+3)** |
| $Ni(L_2)$ | 6 | 0.774**(HOMO)**+2.80*$10^{-3}$**(HOMO-6)**+1.88*$10^{-4}$**(LUMO+2)** | 0.772**(LUMO)**+8.96*$10^{-4}$**(HOMO-3)**+1.20*$10^{-3}$**(LUMO+6)** |
| $Ni(L_2)$-CCSD | 7 | 0.770**(HOMO)**+1.76*$10^{-3}$**(HOMO-6)**+3.90*$10^{-3}$**(LUMO+2)** | 0.759**(LUMO)**+3.20*$10^{-3}$**(HOMO-3)**+8.14*$10^{-3}$**(LUMO+6)** |
| Divinyl-Benzene(CCSD) | 8 | 0.887**(HOMO)**+2.52*$10^{-3}$**(HOMO-1)**+3.64*$10^{-3}$**(LUMO+2)** | 0.884**(LUMO)**+1.65*$10^{-3}$**(HOMO-3)**+9.33*$10^{-4}$**(LUMO+1)** |
| Simplified model of $Ni(L_2)$ | 9 | 0.948**(HOMO)**+2.59*$10^{-4}$**(HOMO-8)**+1.24*$10^{-3}$**(LUMO+2)** | 0.941**(LUMO)+**3.34*$10^{-4}$**(HOMO-5)**+1.36*$10^{-4}$**(HOMO-6)** |
| Simplified model of $Ni(L_2)$(CCSD) | 10 | 0.802**(HOMO)**+8.17*$10^{-4}$**(LUMO)**+3.25*$10^{-3}$**(LUMO+2)** | 0.774**(LUMO)**+7.61*$10^{-4}$**(HOMO)**+3.47*$10^{-4}$**(LUMO+2)** |

Table I. Expansion coefficients of the QPWFs referred to the molecular orbital basis set for the 7 of the 18 molecules examined.

## Supplemental Material of the paper

## ‘Visualizing electron correlation by means of ab-initio scanning tunneling spectroscopy images of single molecules’

by

Dimitrios Toroz, Massimo Rontani, Stefano Corni

**List of contents**

**Discussion S1. Many-body theory of scanning tunneling spectroscopy**

In the following we show that the differential conductance $dI/dV$ originating from the transport resonance between two many-body ground states of a molecule with $N$ and $N-1$ electrons, respectively, as measured by scanning tunneling spectroscopy, is proportional to the h-QPWF square modulus $\left|\varphi(\mathbf{r}_0)\right|^2$, where $\mathbf{r}_0$ is the center of an ideal, non-invasive STM tip that we assume locally spherical. This Appendix recasts Sec. II.A of Ref. 25 as a many-body theory. As remarked in Ref. 25, the evaluation of $\left|\varphi(\mathbf{r}_0)\right|^2$ at $\mathbf{r}_0$, which is possible due to the analytic properties of $\varphi(\mathbf{r})$, is a consequence of the lateral averaging due to the finite tip size.

At zero temperature and small applied bias voltages the differential conductance $dI/dV$ is given by the prediction of first-order time-dependent perturbation theory:

$$\frac{dI}{dV} = \frac{2\pi e^2}{\hbar}\sum_{\xi,i}\left|M_{\xi,i}\right|^2 \delta(\varepsilon_\xi - \mu)\delta(E_i(N) - E_\xi(N-1) - \mu). \qquad \text{(S1)}$$

Here $M_{\xi,i} = \langle \xi | \hat{M} | i \rangle$ is the matrix element of Bardeen's tunneling operator[42,12] $\hat{M}$ between the two many-body states $|i\rangle$ and $|\xi\rangle$, which are both direct products of the tip state times the molecule + substrate state:

$$|i\rangle = |N,i\rangle \otimes |N_{\text{tot}} - N\rangle_{\text{tip}}\ , \qquad \text{(S2)}$$

$$|\xi\rangle = |N-1,\xi\rangle \otimes |N_{\text{tot}} - N + 1\rangle_{\text{tip}}\,. \qquad \text{(S3)}$$

Both states $|i\rangle$ and $|\xi\rangle$ [Eqs. (S2) and (S3), respectively] conserve the total number of electrons, $N_{\text{tot}}$, which is the sum of those in the tip, molecule, and substrate. Furthermore, Dirac's delta appearing in Eq. (S1) ensures that $|i\rangle$ and $|\xi\rangle$ have the same energy. The factor state $|N,i\rangle$ is the many-body $i$th excited state of the $N$-electron molecule + substrate system, the factor state $|N_{\text{tot}} - N\rangle_{\text{tip}}$ is the tip state with $N_{\text{tot}} - N$ non-interacting electrons, the final many-body factor state $|N-1,\xi\rangle$ is obtained by $|N,i\rangle$ by removing an electron and transferring it into the empty single-particle state of the tip whose energy is $\varepsilon_\xi$. Since the tip state is not interacting, its energy variation due to the electron transfer is univocally determined by the difference between the energies of the interacting molecule with $N$ and $N-1$ electrons, respectively $E_i(N)$ and $E_\xi(N-1)$. Note that in Bardeen's approach the factor states are not mutually orthogonal. The explicit form of Bardeen's operator is:[12]

$$\hat{M} = \frac{\hbar^2}{2m}\int\left[\hat{\Psi}^+(\mathbf{r})\frac{\partial\hat{\Psi}(\mathbf{r})}{\partial z} - \frac{\partial\hat{\Psi}^+(\mathbf{r})}{\partial z}\hat{\Psi}(\mathbf{r})\right]\delta(z - z_{\text{vacuum}})d\mathbf{r}\,, \tag{S4}$$

where $m$ is the electron mass, $z$ is the coordinate perpendicular to the substrate (lying in the plane $xy$) which originates from the center of the locally spherical tip and increases as it approaches the substrate, $z_{\text{vacuum}}$ is any coordinate placed in the vacuum region between the tip and the molecule adsorbed on the substrate.

We now write the annihilation Fermi field operator $\hat{\Psi}(\mathbf{r})$ as the sum of tip and molecule + substrate parts, $\hat{\Psi}_T(\mathbf{r})$ and $\hat{\Psi}_S(\mathbf{r})$, respectively:

$$\hat{\Psi}(\mathbf{r}) = \hat{\Psi}_T(\mathbf{r}) + \hat{\Psi}_S(\mathbf{r})\,. \tag{S5}$$

We explicit the tip part $\hat{\Psi}_T(\mathbf{r})$ as

$$\hat{\Psi}_T(\mathbf{r}) = \sum_{\xi}\hat{c}_{\xi}\phi_{\xi}(\mathbf{r})\,, \tag{S6}$$

where the operator $\hat{c}_{\xi}$ destroys an electron occupying the single-particle state $\xi$ of the tip, whose orbital wave function $\phi_{\xi}(\mathbf{r})$ in the region of interest may be approximated as the asymptotic *s*-wave form:

$$\phi_{\xi}(\mathbf{r}) = \frac{1}{\sqrt{\Omega_T}}A_T R\kappa e^{\kappa R}\frac{e^{-\kappa r}}{\kappa r}\,. \tag{S7}$$

Here $\kappa$ is the minimum inverse decay length fixed by the tip work function $W$, $\kappa = \hbar^{-1}(2mW)^{1/2}$, $R$ is the tip radius, $\Omega_T$ is a normalization volume, and $A_T$ is a normalization factor of order one.[25] We also expand the molecule + substrate part $\hat{\Psi}_S(\mathbf{r})$ on the set of two-dimensional plane waves with wave vector $\mathbf{q}$:

$$\hat{\Psi}_S(\mathbf{r}) = \frac{1}{\sqrt{\Omega_S}}\left(\frac{L}{2\pi}\right)^2\int d\mathbf{q}\, e^{i\mathbf{q}\cdot\mathbf{r}} e^{z\left(\kappa^2+q^2\right)^{1/2}}\hat{c}_{\mathbf{q}}\,. \tag{S8}$$

Here $\Omega_S$ is the normalization volume of the molecule, $L^2$ is the area of the substrate on which the molecule is adsorbed, the operator $\hat{c}_{\mathbf{q}}$ destroys an electron filling the wave of vector $\mathbf{q}$ in the $xy$ plane, and we have assumed the same work function as for the tip.

We now evaluate $M_{\xi,i} = \langle \xi | \hat{M} | i \rangle$ by using Eqs. (S2)-(S8) together with the following two-dimensional decomposition for the expression $e^{-\kappa r}/(\kappa r)$ entering Eq. (S7):

$$\frac{e^{-\kappa r}}{\kappa r} = \frac{1}{2\pi}\int d\mathbf{q} \frac{1}{\kappa\left(\kappa^2 + q^2\right)^{1/2}} e^{-\left(\kappa^2+q^2\right)^{1/2}|z|} e^{i\mathbf{q}\cdot\mathbf{r}} . \tag{S9}$$

The result is

$$M_{\xi,i} = \frac{\hbar^2}{2m}\frac{4\pi}{\sqrt{\Omega_T}}\kappa^{-1}A_T R\kappa e^{\kappa R}\frac{1}{\sqrt{\Omega_S}}\left(\frac{L}{2\pi}\right)^2 \int d\mathbf{q} \langle N-1, \xi | \hat{c}_{\mathbf{q}} | N, i \rangle$$

$$= \frac{\hbar^2}{2m}\frac{4\pi}{\sqrt{\Omega_T}}\kappa^{-1}A_T R\kappa e^{\kappa R} \langle N-1, \xi | \hat{\Psi}_S(0) | N, i \rangle . \tag{S10}$$

Therefore, if only the ground state $|N\rangle$ of the adsorbed molecule is involved in the sum over $i$ of Eq. (S1) –a typical situation for low-temperature STS- we obtain that $dI/dV$ is directly proportional to the spectral density of the molecule + substrate system evaluated at the chemical potential and frame origin, which is the center of the tip $\mathbf{r}_0$:

$$\frac{dI}{dV} \propto N(\mathbf{r}_0, \mu/\hbar) . \tag{S11}$$

Eventually, if we are concerned only with the tunneling event connecting $|N\rangle$ and $|N-1\rangle$:

$$\frac{dI}{dV} \propto |\varphi(\mathbf{r}_0)|^2 . \tag{S12}$$

QED.

We conclude this Discussion by recalling that the above approach, based on the perturbative treatment of the tip-molecule coupling, is unable to treat many-body phenomena such as the Kondo effect. In the latter case a many-body state coherent over the whole tip + molecule + substrate system sets in requiring the knowledge of the global spectral density [cf. M. Rontani, Phys. Rev. Lett. **97,** 076801 (2006); Phys. Rev. B **82,** 045310 (2010)].

**Discussion S2. Supplementary results**

From the halogen series derivatives of paradivinylbenzene we observe that for the fluoro-divinylbenzene (Fig. S1) the discrepancies remain minimal and the images obtained are very similar to those predicted for divinylbenzene. Larger modifications, although still not observable, occur for the bromine (Fig. S3), chlorine (Fig. S2), and iodine (Fig. S4) substituents, especially as far as the orbital lobes of the e-QPWF are concerned. For example, in the e-QPWF-ZERO image for bromo-divinylbenzene (Fig. S3) the orbital lobe which appears at the positions 1 and 4 of the benzene ring is denser in comparison with the other halogen derivatives of divinylbenzene. This explains the depletion of weight in the e-QPWF with respect to LUMO-HF, as confirmed by the SUBL image, which displays a larger weight in comparison to that obtained for the other derivatives. We note that in the h-QPWF-ZERO image of Fig. S3 orbital lobes do not cover the benzene ring as it has been observed for the other substituents but they interfere destructively in position 1. We also note that in iodo-divilynbenzene (Fig. S4) the discrepancies found are smaller than those for bromo-divinylbenzene (Fig. S3). This is immediately seen in the SUBH and SUBL images where the orbital lobes are almost invisible. However, the two images are not strictly comparable. In fact, the iodine derivative calculations have been performed with the LANL2DZ basis set since it was not feasible to exploit the cc-PVDZ basis set used for the other substituents.

The results obtained for the cyano, the hydroxyl (enol tautomer), the thiol, and the nitro groups are similar to those observed with the halogen series. The effect of correlation observed in the divinylbenzene with the cyanide group (Fig. S5) is similar to that observed for the bromine derivative (Fig. S3). The images obtained for the divinylbenzene-thiol derivative (Fig. S6) are almost identical with the results obtained for the chlorine atom derivative (Fig. S2). Analogously, the images obtained for the divinylbenzene derivative including the hydroxyl group (Fig. S7) are very similar to those for the fluorine derivative (Fig. S1).

The HOMO and LUMO and the corresponding QPWFs images for nitrobenzene were also calculated by employing the larger basis set cc-PVTZ (Figure S8). The plots of Fig. S8 confirm the scenario illustrated for nitrobenzene (cf. Fig. 3), even if the orbital lobes of the h-QPWF and the e-QPWF are additionally reduced. In fact, SUBH and SUBL plots of Fig. S8 clearly show that the orbital modification predicted with the cc-PVTZ basis set is stronger than with the smaller cc-PVDZ set (cf. Fig. 3). This trend is expected, since the larger (more flexible) basis set allows for more excitations and thus a better description of correlation, due to the variational character of the CI technique.

Figure S9 presents the images of the HOMO and LUMO and the corresponding QPWFs of dinitrobenzene. The wave function plots resemble those of nitrobenzene. In particular, HF frontier orbitals and QPWFs are essentially indistinguishable. Additionally, QPWF-ZERO images are similar to those of nitrobenzene. The next molecule we have tested is tri-nitrobenzene. At the HF level there are two degenerate frontier states for both HOMO and LUMO, as shown in Fig. S10. The discrepancies between the HF orbitals and the QPWFs are still minimal, with regards to both topology and shape of the orbitals. In the h-QPWF-ZERO and e-QPWF-ZERO images of Fig. S10, in contrast to those of nitro-benzene (Fig. 3) and di-nitrobenzene (Fig. S9), there are four orbital lobes within the benzene ring due to the presence of the three nitro groups.

Eventually, in Fig. S11 a larger molecule containing two phenyl and two nitro groups (4,4 dinitro biphenyl) has been considered. The results are similar to those for nitrobenzene and dinitrobenzene. The only direct discrepancy identified, in comparison to nitrobenzene and dinitrobenzene, is on the orbital lobes that define the oxygen atoms of the two nitro groups. In fact, these lobes in the SUBH and SUBL are denser, which shows that the presence of two phenyl groups increases the effect of correlation.

Figure S12 presents the the HOMO and LUMO and the corresponding QPWFs of naphthalene. The QPWFs obtained for naphthalene are very similar to the HF frontier orbitals (Fig. S12). The effects of correlation may be observed only in QPWF-ZERO images, taking place within the fused rings.

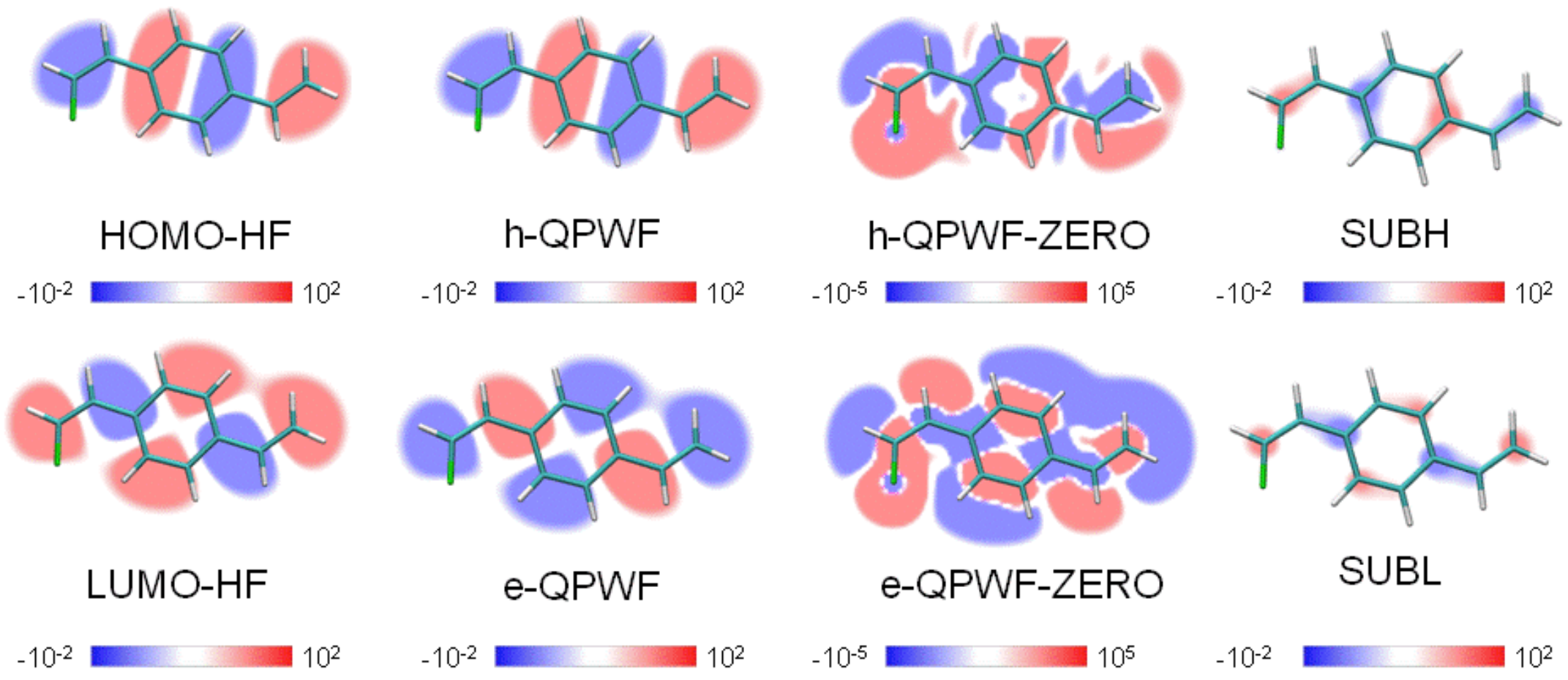

**Figure S1.** Slices of frontier orbitals (i.e., HOMO and LUMO), QPWFs (h-QPWF and e-QPWF), QPWFs omitting the contribution on the HOMO and LUMO orbital (h-QPWF-ZERO and e-QPWF-ZERO), and result of the subtraction between the HOMO and h-QPWF-ZERO (SUBH) and the LUMO with e-QPWF-ZERO (SUBL) of fluoro-divinylbenzene.

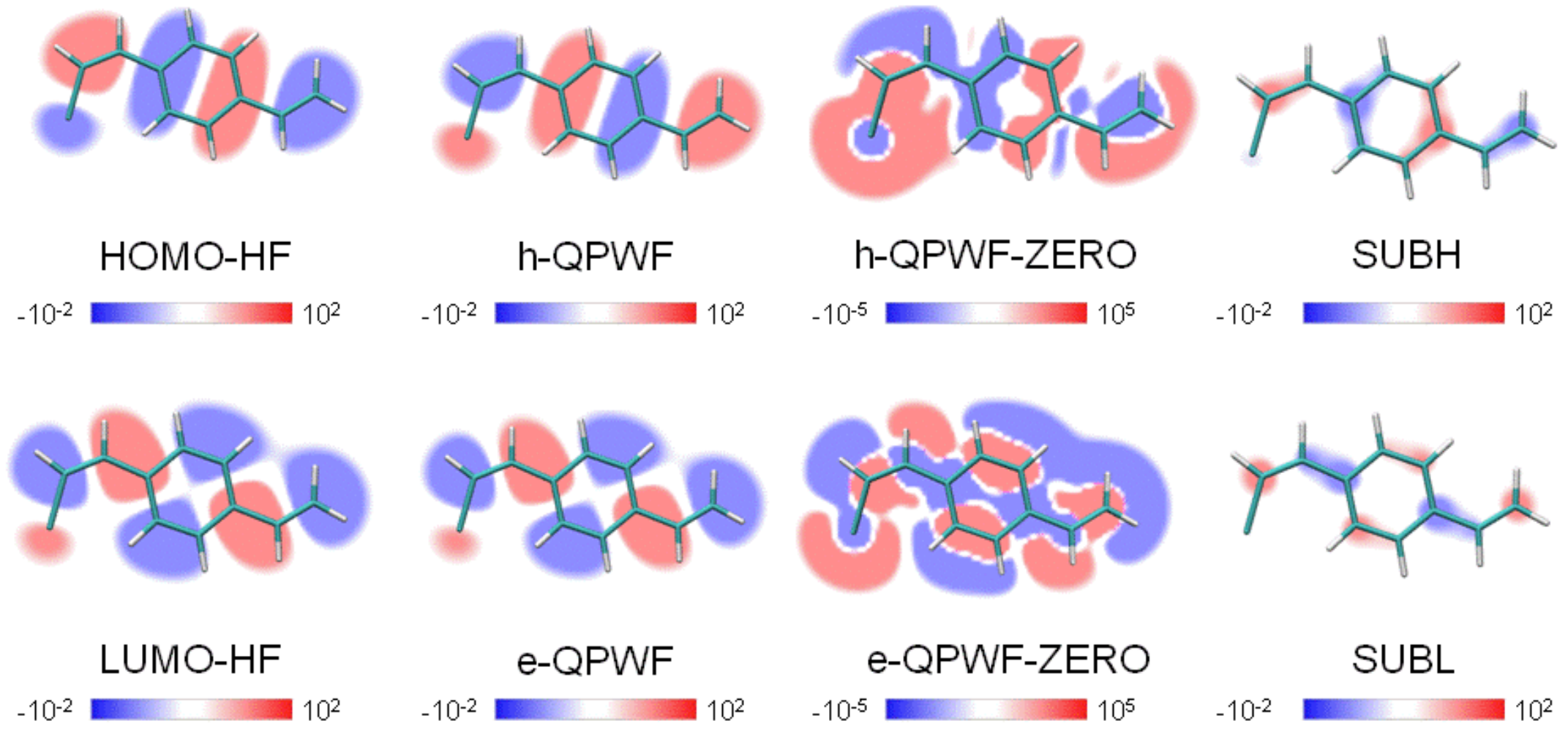

**Figure S2.** Slices of frontier orbitals (i.e., HOMO and LUMO), QPWFs (h-QPWF and e-QPWF), QPWFs omitting the contribution on the HOMO and LUMO orbital (h-QPWF-ZERO and e-QPWF-ZERO), and result of the subtraction between the HOMO and h-QPWF-ZERO (SUBH) and the LUMO with e-QPWF-ZERO (SUBL) of chloro-divinylbenzene.

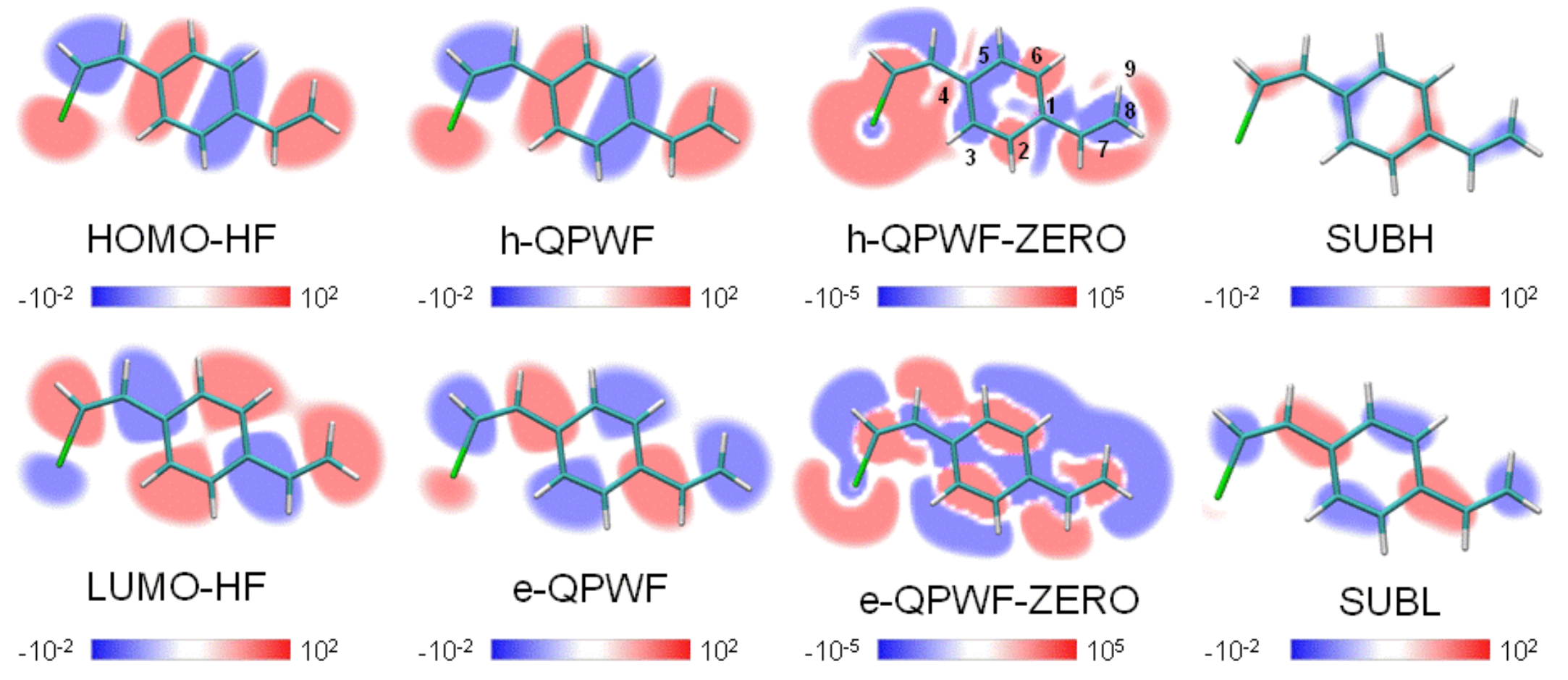

**Figure S3.** Slices of degenerate frontier orbitals (i.e., HOMO and LUMO), QPWFs (h-QPWF and e-QPWF), QPWFs omitting the contribution on the HOMO and LUMO orbital (h-QPWF-ZERO and e-QPWF-ZERO), and result of the subtraction between the HOMO and h-QPWF-ZERO (SUBH) and the LUMO with e-QPWF-ZERO (SUBL) of bromo-divinylbenzene.

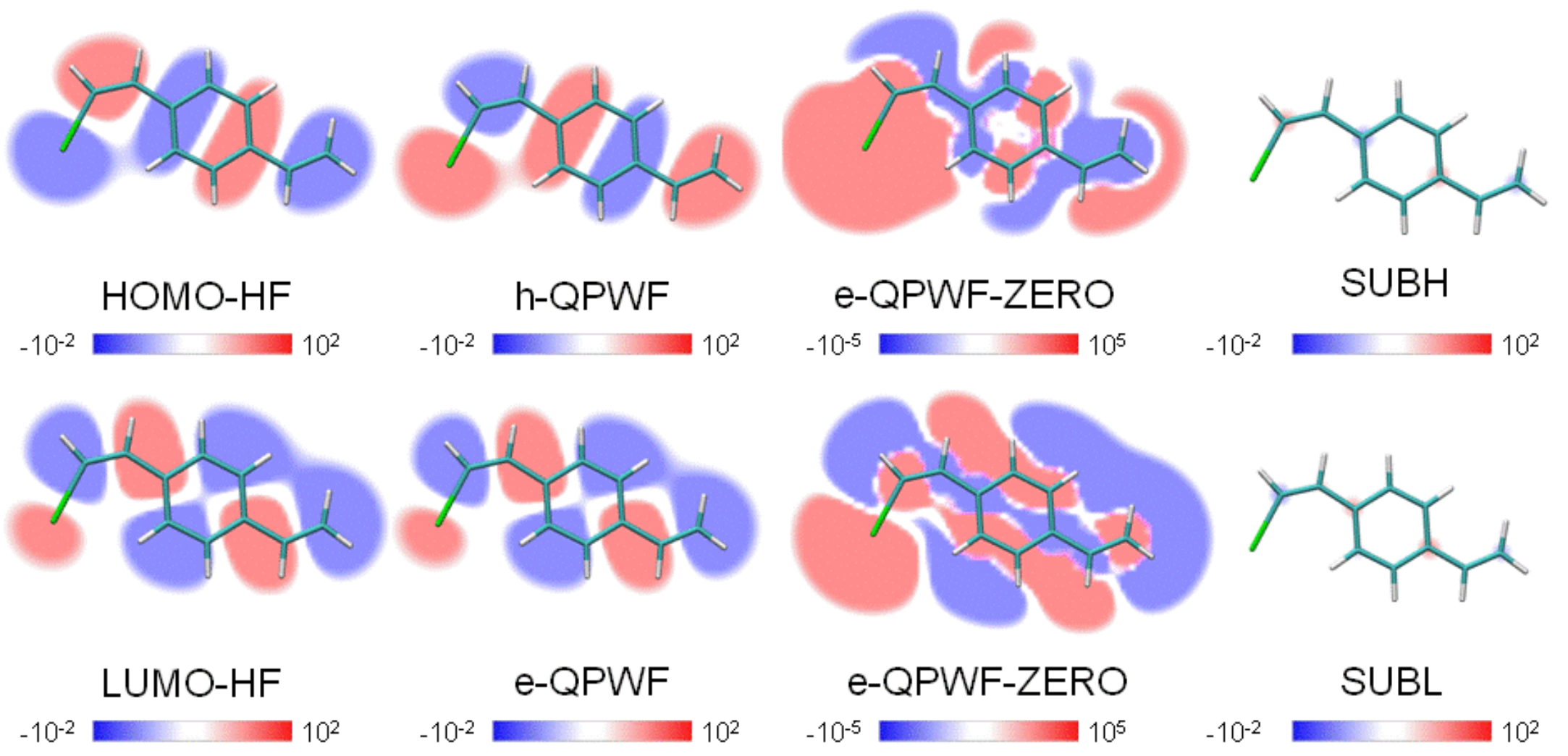

**Figure S4.** Slices of frontier orbitals (i.e., HOMO and LUMO), QPWFs (h-QPWF and e-QPWF), QPWFs omitting the contribution on the HOMO and LUMO orbital (h-QPWF-ZERO and e-QPWF-ZERO), and result of the subtraction between the HOMO and h-QPWF-ZERO (SUBH) and the LUMO with e-QPWF-ZERO (SUBL) of iodo-divinylbenzene.

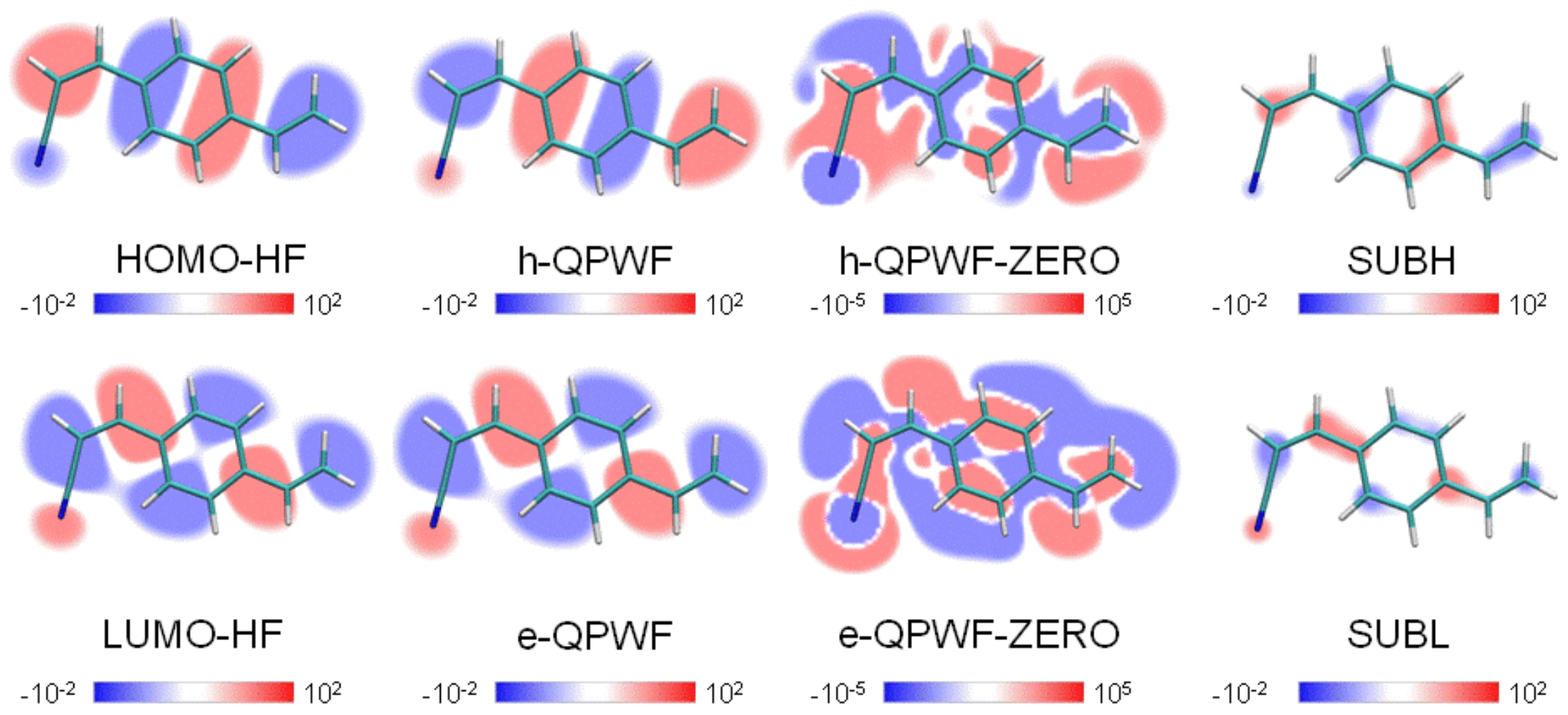

**Figure S5.** Slices of frontier orbitals (i.e., HOMO and LUMO), QPWFs (h-QPWF and e-QPWF), QPWFs omitting the contribution on the HOMO and LUMO orbital (h-QPWF-ZERO and e-QPWF-ZERO), and result of the subtraction between the HOMO and h-QPWF-ZERO (SUBH) and the LUMO with e-QPWF-ZERO (SUBL) of cyano-divinylbenzene.

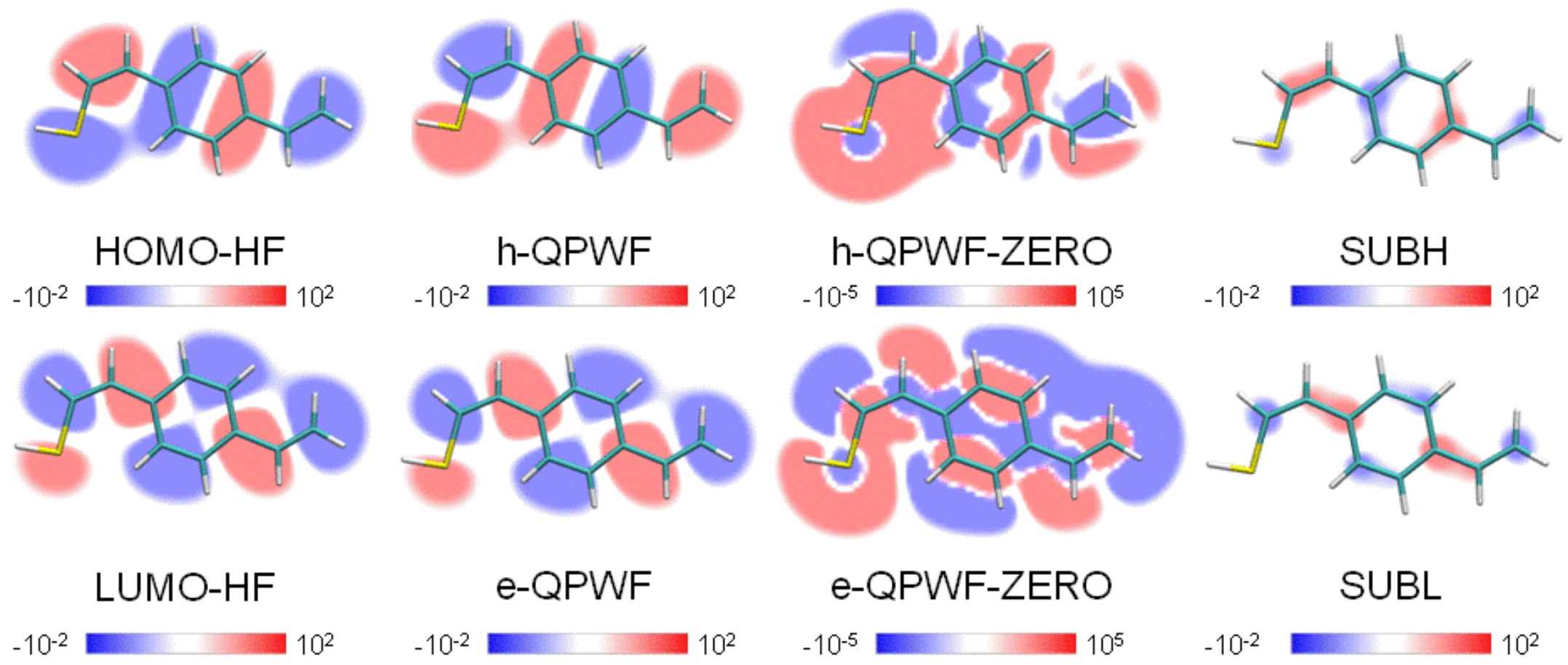

**Figure S6.** Slices of frontier orbitals (i.e., HOMO and LUMO), QPWFs (h-QPWF and e-QPWF), QPWFs omitting the contribution on the HOMO and LUMO orbital (h-QPWF-ZERO and e-QPWF-ZERO), and result of the subtraction between the HOMO and h-QPWF-ZERO (SUBH) and the LUMO with e-QPWF-ZERO (SUBL) of thiol-divinylbenzene.

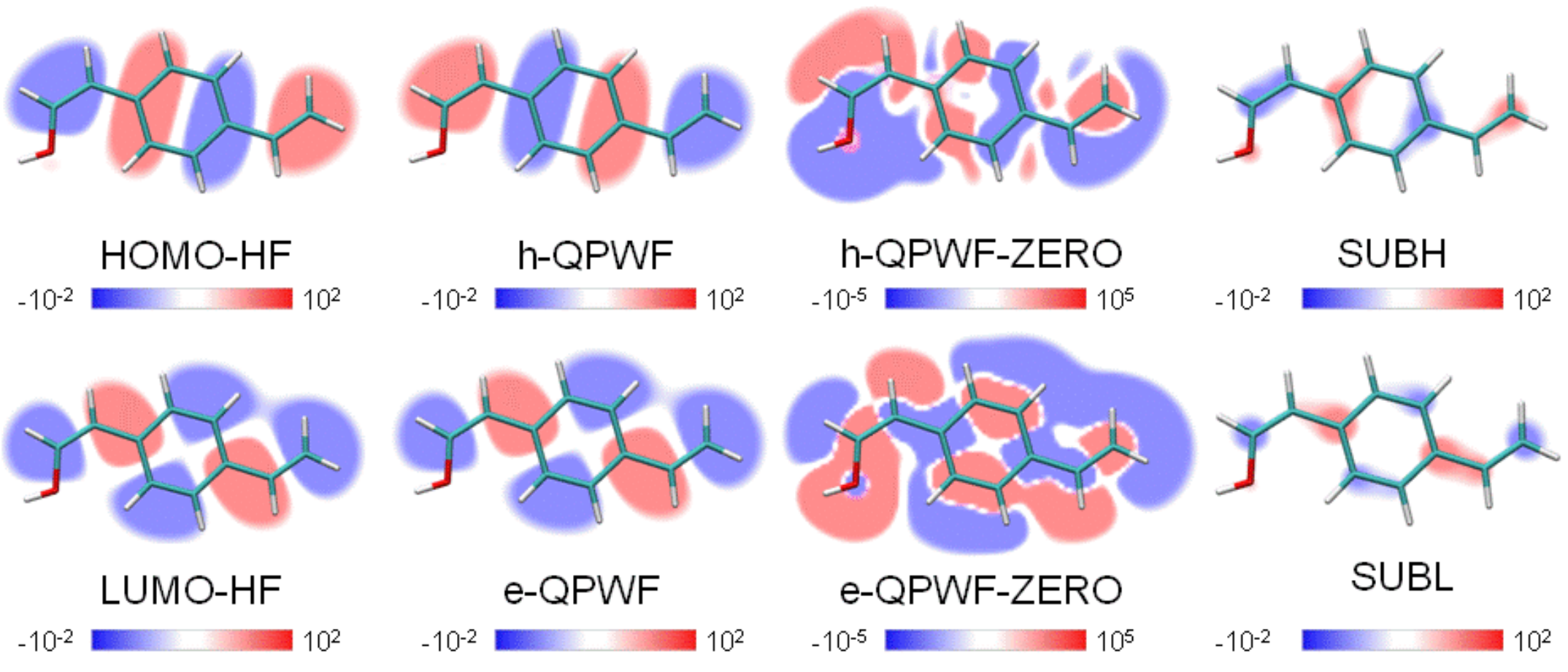

**Figure S7.** Slices of frontier orbitals (i.e., HOMO and LUMO), QPWFs (h-QPWF and e-QPWF), QPWFs omitting the contribution on the HOMO and LUMO orbital (h-QPWF-ZERO and e-QPWF-ZERO), and result of the subtraction between the HOMO and h-QPWF-ZERO (SUBH) and the LUMO with e-QPWF-ZERO (SUBL) of hydroxy-divinylbenzene.

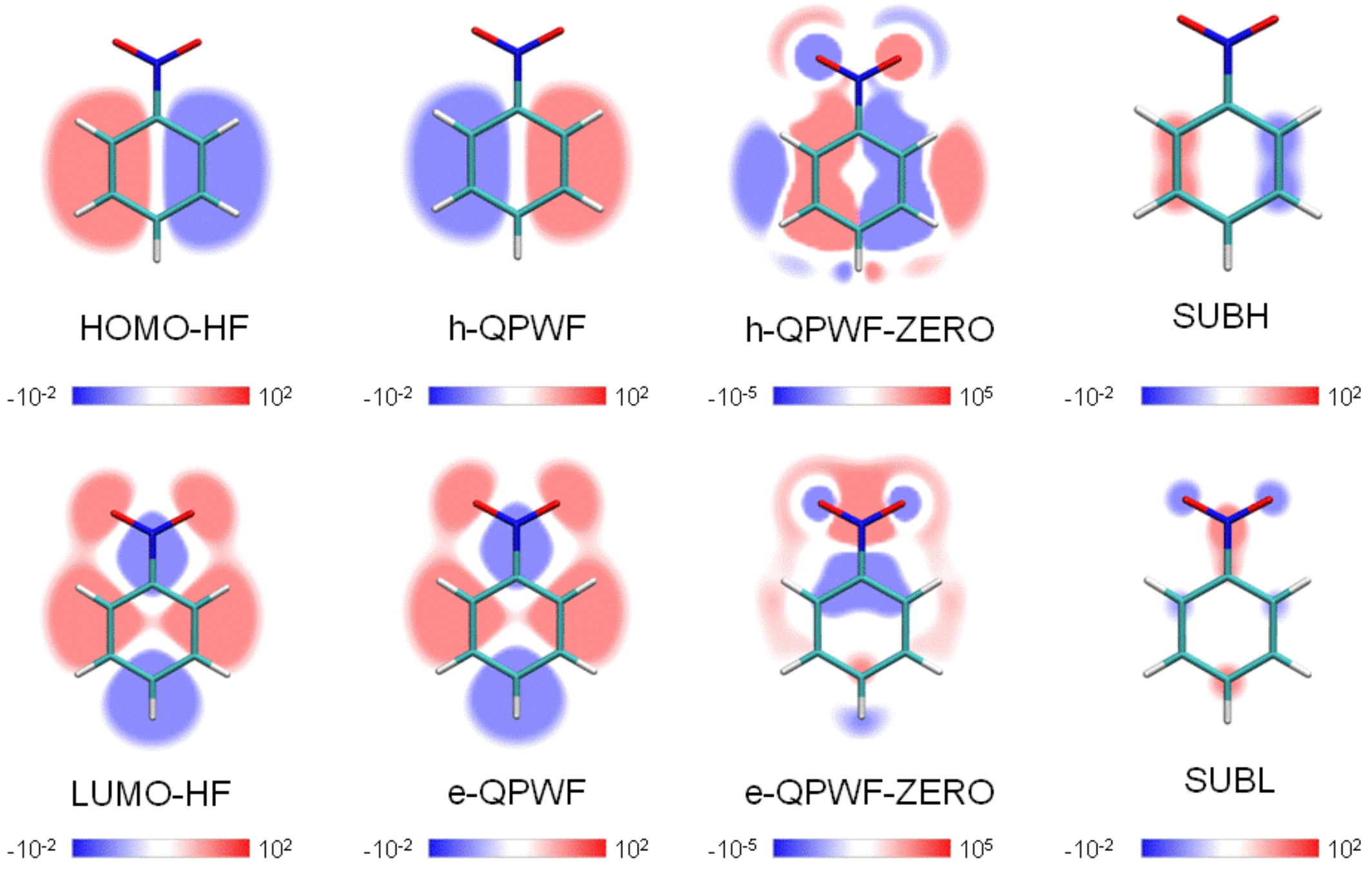

**Figure S8.** Slices of frontier orbitals (i.e., HOMO and LUMO), QPWFs (h-QPWF and e-QPWF), QPWFs omitting the contribution on the HOMO and LUMO orbital (h-QPWF-ZERO and e-QPWF-ZERO), and result of the subtraction between the HOMO and h-QPWF-ZERO (SUBH) and the LUMO with e-QPWF-ZERO (SUBL) of nitrobenzene using the cc-PVTZ basis set.

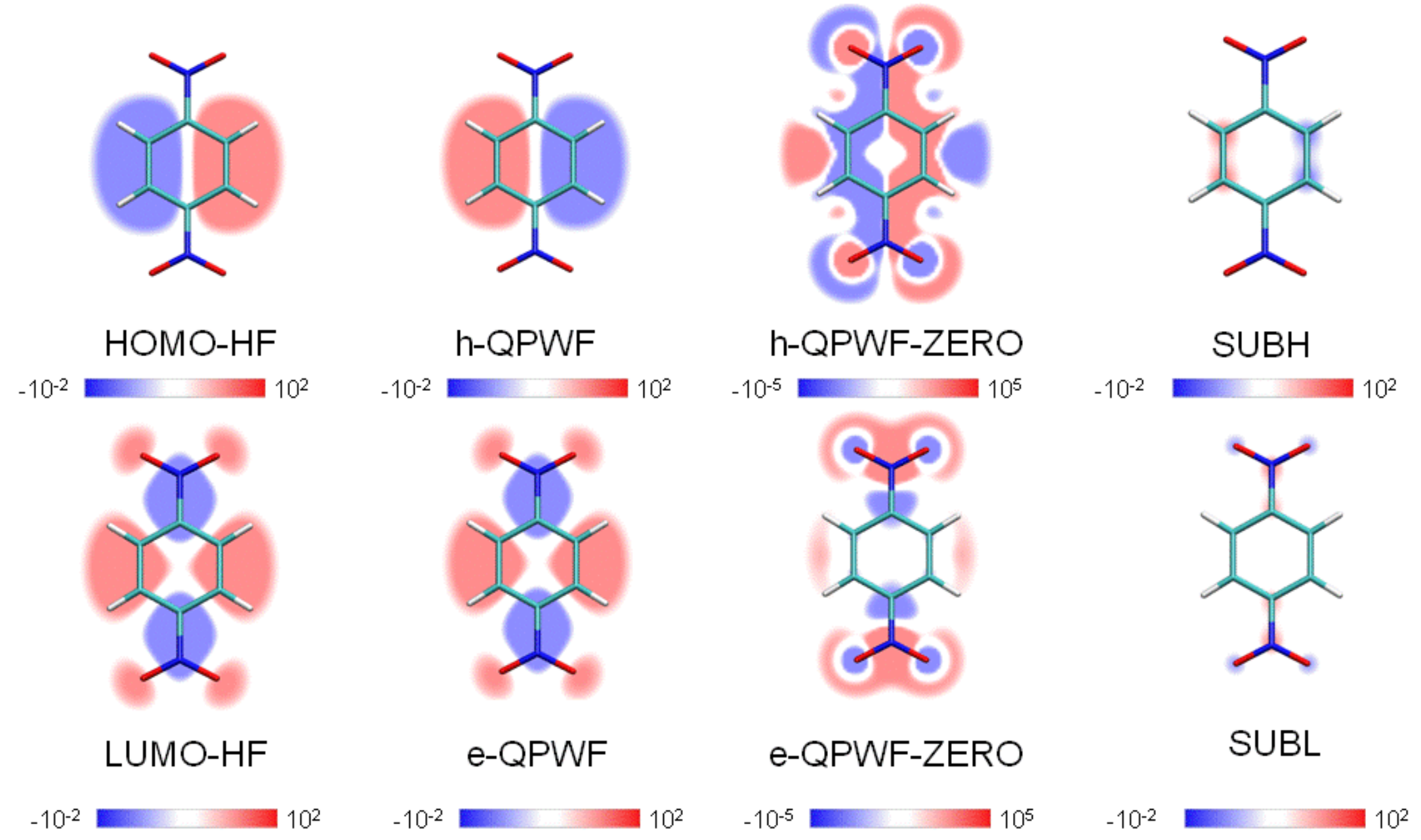

**Figure S9.** Slices of frontier orbitals (i.e., HOMO and LUMO), QPWFs (h-QPWF and e-QPWF), QPWFs omitting the contribution on the HOMO and LUMO orbital (h-QPWF-ZERO and e-QPWF-ZERO), and result of the subtraction between the HOMO and h-QPWF-ZERO (SUBH) and the LUMO with e-QPWF-ZERO (SUBL) of di-nitrobenzene.

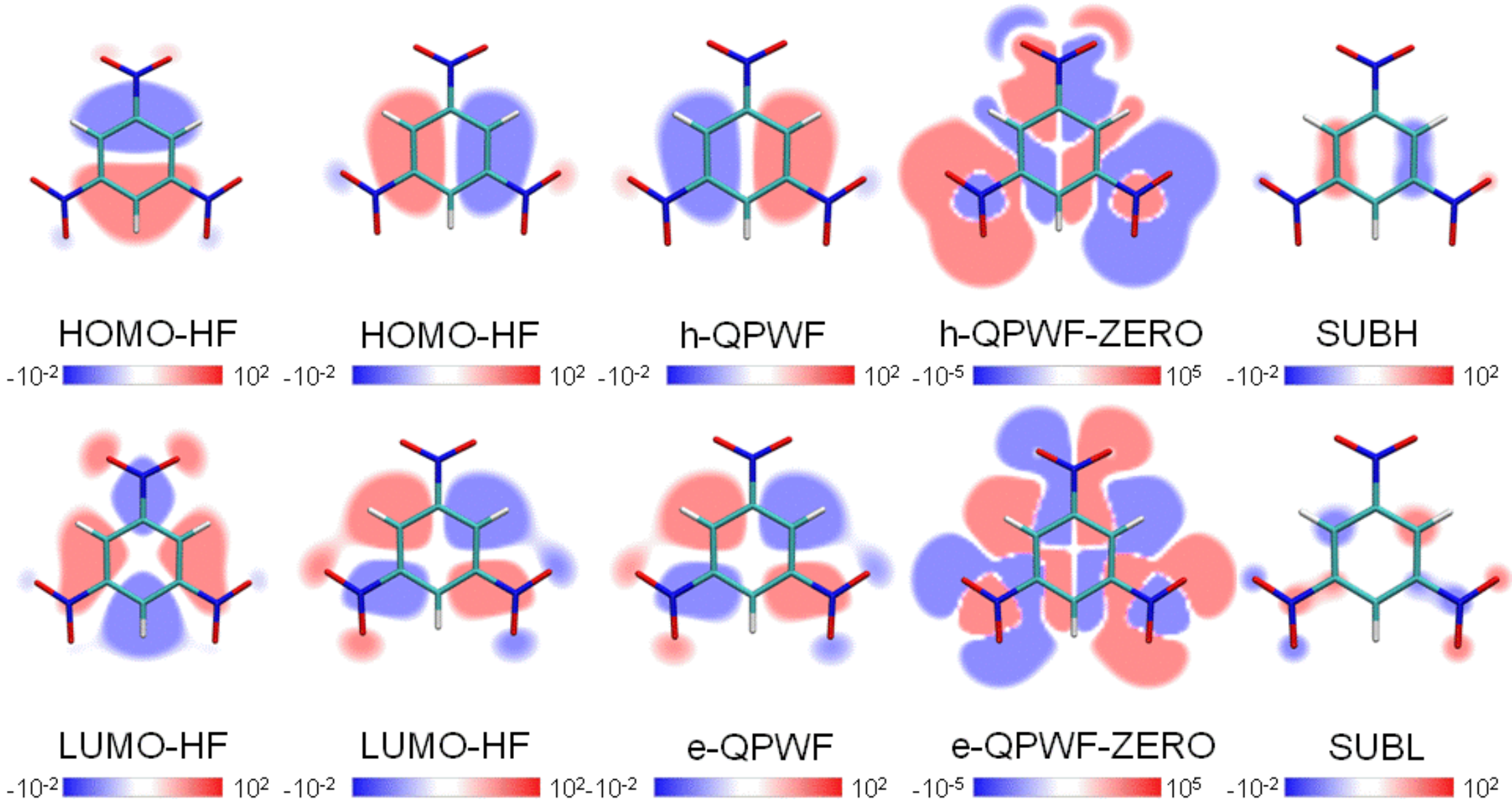

**Figure S10.** Slices of degenerate frontier orbitals (i.e., HOMO and LUMO), QPWFs (h-QPWF and e-QPWF), QPWFs omitting the contribution on the HOMO and LUMO orbital (h-QPWF-ZERO and e-QPWF-ZERO), and result of the subtraction between the HOMO and h-QPWF-ZERO (SUBH) and the LUMO with e-QPWF-ZERO (SUBL) of tri-nitrobenzene.

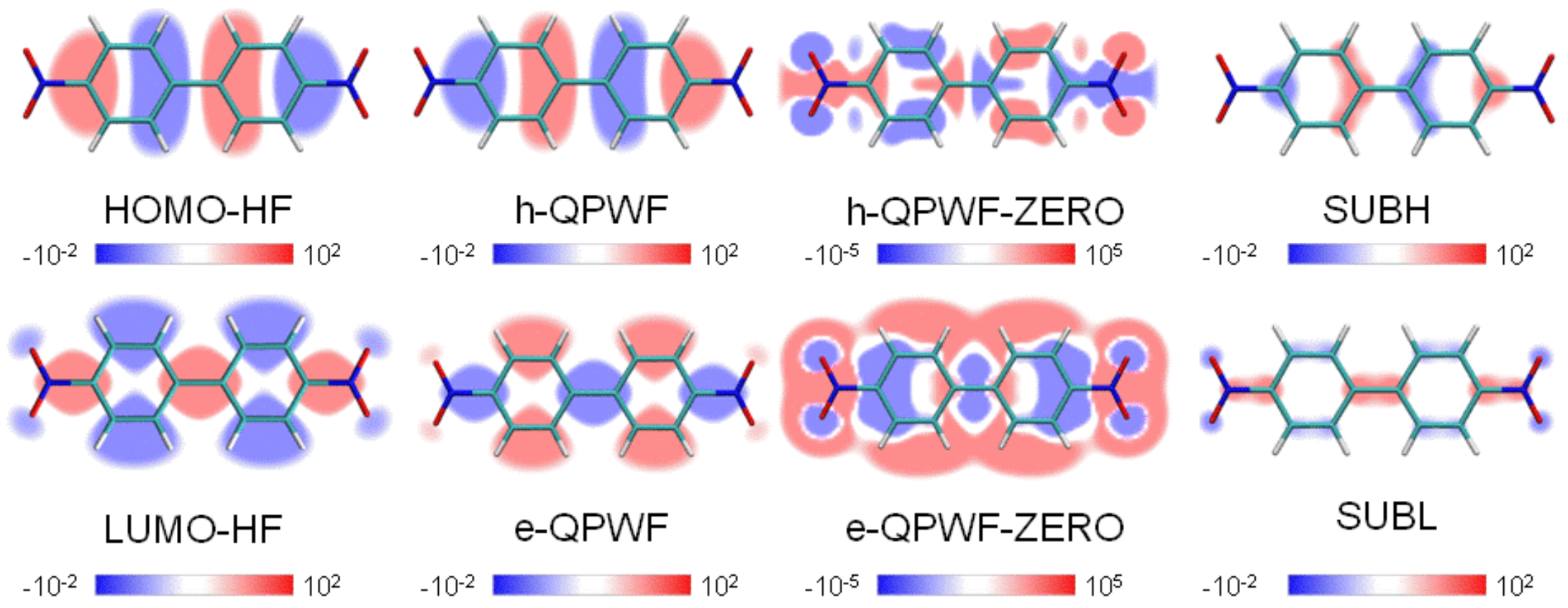

**Figure S11.** Slices of frontier orbitals (i.e., HOMO and LUMO), QPWFs (h-QPWF and e-QPWF), QPWFs omitting the contribution on the HOMO and LUMO orbital (h-QPWF-ZERO and e-QPWF-ZERO), and result of the subtraction between the HOMO and h-QPWF-ZERO (SUBH) and the LUMO with e-QPWF-ZERO (SUBL) of 4,4-dinitro-1,1-biphenyl.

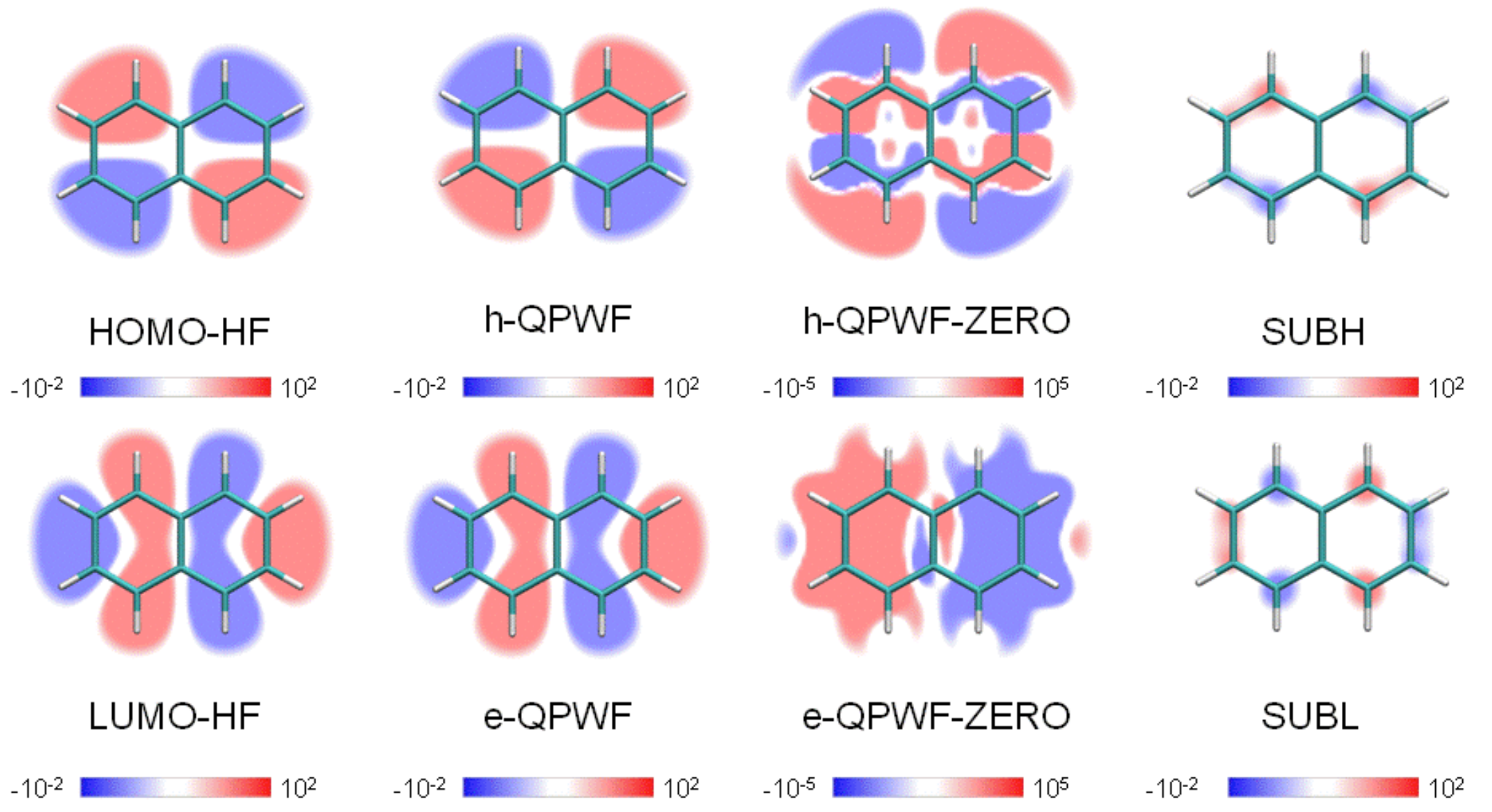

**Figure S12.** Slices of frontier orbitals (i.e., HOMO and LUMO), QPWFs (h-QPWF and e-QPWF), QPWFs omitting the contribution on the HOMO and LUMO orbital (h-QPWF-ZERO and e-QPWF-ZERO), and result of the subtraction between the HOMO and h-QPWF-ZERO (SUBH) and the LUMO with e-QPWF-ZERO (SUBL) of naphthalene.

| Molecule | Figure | h-QPWF | e-QPWF |
|---|---|---|---|
| Fluoro-Divinyl-Benzene | S1 | 0.896**(HOMO)**+1.98*$10^{-4}$**(LUMO+2)** | 0.893**(LUMO)**+1.73*$10^{-4}$**(LUMO+1)**+1.58*$10^{-4}$**(LUMO+2)** |
| Chloro-Divinyl-Benzene | S2 | 0.888**(HOMO)**+1.04*$10^{-4}$**(HOMO-3)**+1.61*$10^{-4}$**(LUMO)** | 0.885**(LUMO)**+1.86*$10^{-4}$**(LUMO+1**)+1.09*$10^{-4}$**(LUMO+2)** |
| Bromo-Divinyl-Benzene | S3 | 0.887**(HOMO)**+1.77*$10^{-4}$**(HOMO-2)**+1.60*$10^{-4}$**(LUMO)** | 0.884**(LUMO)**+1.87*$10^{-4}$**(LUMO+1)**+2.22*$10^{-4}$**(HOMO-5)** |
| Iodo-Divinyl-Benzene | S4 | 0.934**(HOMO)**+7.65*$10^{-4}$**(HOMO-1)**+1.69*$10^{-4}$**(LUMO)** | 0.932**(LUMO)**+2.96*$10^{-4}$**(HOMO)**+5.39*$10^{-4}$**(LUMO+2)** |
| Cyano-Divinyl-Benzene | S5 | 0.882**(HOMO)**+1.20*$10^{-4}$**(HOMO-3)**+4.47*$10^{-4}$**(LUMO)** | 0.879**(LUMO)**+2.42*$10^{-4}$**(HOMO)**+1.90*$10^{-4}$**(LUMO+1)** |
| Thiol-Divinyl-Benzene | S6 | 0.885(**HOMO)**+2.04*$10^{-4}$**(HOMO-1)**+3.32*$10^{-4}$**(HOMO-2)** | 0.882**(LUMO)**+1.61*$10^{-4}$**(LUMO+1)**+1.72*$10^{-4}$**(LUMO+3)** |
| Hydroxy-Divinyl-Benzene | S7 | 0.890**(HOMO)**+1.06*$10^{-4}$**(HOMO-2)**+2.20*$10^{-4}$**(LUMO+3)** | 0.887**(LUMO)**+1.46*$10^{-4}$**(HOMO)**+1.78*$10^{-4}$**(LUMO+1)** |
| NItrobenzene-cc-PVTZ | S8 | 0.915**(HOMO)**+2.07*$10^{-4}$**(HOMO-2)**+1.70*$10^{-4}$**(LUMO+1)** | 0.909**(LUMO)**+8.44*$10^{-4}$**(HOMO-1)**+2.37*$10^{-3}$**(LUMO+3)** |
| Dinitrobenzene | S9 | 0.937**(HOMO)**+2.47*$10^{-4}$**(HOMO-3)** | 0.932(**LUMO)**+2.42*$10^{-3}$**(LUMO+3)** |
| Trinitrobenzene | S10 | 0.895**(HOMO)**+1.43*$10^{-4}$**(HOMO-3)**+1.07*$10^{-3}$**(LUMO)** | 0.892**(LUMO)**+1.82*$10^{-4}$**(HOMO)**+1.97*$10^{-3}$**(LUMO+4)** |
| 4,4-dinitro-1.1-biphenyl | S11 | 0.872**(HOMO)**+7.14*$10^{-4}$**(LUMO+1)** | 0.869**(LUMO)**+2.45*$10^{-4}$**(HOMO-3)**+1.33*$10^{-3}$**(LUMO+4)** |
| Naphthalene | S12 | 0.913**(HOMO)**+8.13*$10^{-4}$**(LUMO+10)** | 0.899**(LUMO)**+1.35*$10^{-3}$**(HOMO-3)** |

**Table SI.** Expansion coefficients of the QPWFs referred to the molecular orbital basis set for the divinylbenzene derivatives, Nitrobenzene using the cc-PVTZ basis set, Dinitrobenzene, Trinitrobenzene, 4,4-dinitro-1.1-biphenyl and naphthalene.